\documentclass[onecolumn,11pt]{IEEEtran}
\usepackage{amsmath,amssymb}
\usepackage{graphicx}
\usepackage{setspace}
\usepackage{pgf}
\usepackage{tikz}
\usepackage{cite}
\usepackage{pgffor}
\usepackage{pgfplots}
\usepackage{pgfplotstable}
\usetikzlibrary{arrows,shapes,positioning,plotmarks,fit}
\usepgflibrary{shapes.geometric}
\newcommand{\norm}[1]{\left\|{#1}\right\|}
\def\Hbf{\mathbf{H}}
\def\Gbf{\mathbf{G}}
\def\Fbf{\mathbf{F}}

\def\xbf{\mathbf{x}}

\def\cbf{\mathbf{c}}
\def\tbf{\mathbf{t}}
\def\fbf{\mathbf{f}}
\def\sbf{\mathbf{s}}

\def\wbf{\mathbf{w}}
\def\vbf{\mathbf{v}}
\def\ybf{\mathbf{y}}
\def\zbf{\mathbf{z}}
\def\nbf{\mathbf{n}}

\def\hbf{\mathbf{h}}

\def\zbf{\mathbf{z}}
\def\Vcal{\mathcal{V}}
\def\Ccal{\mathcal{C}}

\def\Lcal{\mathcal{L}}

\def\Re{\mathbb{R}}

\newtheorem{theorem}{Theorem}

\newtheorem{corollary}{Corollary}

\newtheorem{lemma}{Lemma}
\newtheorem{example}{Example}

\makeatletter
\def\mod#1{\  \mathrm{mod} \ #1}
\makeatother

\title{Cooperative Compute-and-Forward}
\author{Matthew~Nokleby, {\em Student Member, IEEE,} and Behnaam~Aazhang, {\em Fellow, IEEE} \thanks{email: \{nokleby, aaz\}@rice.edu. Matthew Nokleby and Behnaam Aazhang are with Rice University, Houston, TX. Behnaam Aazhang is also the Finnish Academy Distinguished Professor (FiDiPro) at the Center for Wireless Communication, University of Oulu, Finland. This work is supported in part by Renesas Mobile and the National Science Foundation. This work was presented in part at the IEEE Symposium on Information Theory, St. Petersburg, Russia, August 2011, and the IEEE Information Theory Workshop, Paraty, Brazil, October 2011.}}

\begin{document}
\maketitle
\onehalfspacing

\begin{abstract}
We examine the benefits of user cooperation under compute-and-forward. Much like in network coding, receivers in a compute-and-forward network recover finite-field linear combinations of transmitters' messages. Recovery is enabled by linear codes: transmitters map messages to a linear codebook, and receivers attempt to decode the incoming superposition of signals to an integer combination of codewords. However, the achievable computation rates are low if channel gains do not correspond to a suitable linear combination. In response to this challenge, we propose a cooperative approach to compute-and-forward. We devise a lattice-coding approach to block Markov encoding with which we construct a decode-and-forward style computation strategy. Transmitters broadcast lattice codewords, decode each other's messages, and then cooperatively transmit resolution information to aid receivers in decoding the integer combinations. Using our strategy, we show that cooperation offers a significant improvement both in the achievable computation rate and in the diversity-multiplexing tradeoff.
\end{abstract}

\section{Introduction}
Interference is the primary obstacle to communications over wireless networks. Due to the broadcast nature of the wireless medium, a transmitter's signal arrives not only at its intended receiver(s), but also at any terminal in the vicinity. This fact has proven to be a formidable challenge. Despite decades of study and a plethora of sophisticated techniques, the capacity of even the two-user interference channel remains unknown in general.

Most approaches to interference entail the minimization of its effects. A special case of the Han-Kobayashi scheme \cite{han:IT81}, in which receivers decode a portion of the interference, was recently shown to achieve rates within one bit of the capacity region of the two-user interference channel \cite{etkin:IT08}. Interference alignment, in which interfering signals are made to lie in a low-dimensional subspace by means of multiple antennas, changing channel conditions, or signal-scale techniques, has been shown to provide the optimal {\em degrees of freedom} of the interference channel: for large signal-to-noise ratios, each transmitter can achieve approximately half the rate possible in the absence of interference \cite{maddah-ali:IT08,cadambe:IT08,nazer:IT09,motahari:IT09}. In a somewhat similar approach, lattice codes are used in the many-to-one interference channel to align interference at the signal scale, allowing the receiver to decode the interference as though it came from a single receiver \cite{bresler:IT10}. In each of these strategies, the goal is to minimize the effective interference seen by each receiver.

{\em Compute-and-forward} \cite{nazer:IT11}, also known as {\em physical-layer network coding} \cite{nazer:IEEE11} is an innovative technique which exploits, rather than eliminates, interference. Under compute-and-forward, receivers decode finite-field linear combinations of transmitter's messages instead of messages themselves. If enough linearly independent combinations are recovered, the individual messages can be recovered further ``downstream'' in the network. In \cite{nazer:IT11} such decoding is enabled by lattice codes. Transmitters send out lattice codewords, noisy linear combinations of which arrive at the receivers. Each receiver decodes the incoming signal to an integer combination of the lattice codewords corresponding to the desired finite-field linear combination. Since an integer combination of lattice points is itself a lattice point, it can be decoded almost as though it were a single incoming signal. Linear combinations of messages therefore are often easier to decode than individual messages. 

The strategy proposed in \cite{nazer:IT11} requires a correspondence between the channel gains and the desired integer combinations. If the channels do not produce suitable linear combinations of transmitters' signals, the receivers cannot easily recover suitable integer combinations of the lattice points. Several solutions to this challenge have been proposed. {\em Integer-forcing receivers} \cite{zhan:ISIT09,zhan:VTC10}, in which linear receivers are chosen to induce integer-valued equivalent channels, were developed for compute-and-forward over multiple-input multiple-output (MIMO) channels. In \cite{niesen:ISIT11}, a number-theoretic approach was developed to address this problem in the high-SNR regime. Using techniques from Diophantine approximations, an encoding strategy was proposed that achieves the full degrees of freedom.

We take a different tack. Our approach is based on the observation that, if transmitters were able to encode their messages jointly, compute-and-forward would reduce to a multiple-antenna broadcast channel, the capacity of which is known \cite{weingarten:IT06}. While {\em perfect} cooperation is infeasible, users can cooperate {\em partially} by exploiting another consequence of the broadcast nature: transmitters can overhear each other's signals and jointly encode portions of their messages. The aim of this paper therefore is to examine the extent to which user cooperation can improve the performance of compute-and-forward.

Our main contribution is a cooperative strategy for compute-and-forward. We develop a lattice-coding instantiation of block Markov encoding by decomposing the lattice codebook into two linearly independent, lower-rate constituent codes, called the {\em resolution} codebook and the {\em vestigial} codebook. Transmitters broadcast lattice codewords, after which they decode the codewords of other transmitters. They then transmit cooperatively the resolution codewords corresponding to the linear combinations desired at the receivers. Receivers employ a variant of sliding-window decoding tailored to our lattice decomposition. They decode the resolution information and subtract it from the original signal; they then need only to decode the vestigial component of the desired sum of lattice points. This strategy allows an improvement in computation rate due to two factors. First, since cooperating transmitters decode others' messages, they can jointly encode portions of the linear combinations directly, relaxing the need for receivers to recover the messages from separately-encoded signals. Second, the jointly encoded signals combine coherently at receivers, resulting in a beamforming gain.


We also present a high-SNR analysis of our approach. User cooperation naturally lends itself to diversity gains, and we show that our approach indeed increases the diversity order under slow Rayleigh fading. We focus on the case of a single receiver. We prove inner bounds on the diversity-multiplexing tradeoff (DMT) using a combination of cooperative random coding techniques and our cooperative lattice strategy. Random coding achieves full diversity but performs poorly at high multiplexing gains, whereas lattice coding falls short of full diversity but maintains performance at high multiplexing gains. Between the two coding strategies we establish an achievable DMT whose corner points match the upper bound of a multiple-antenna, single output (MISO) system.


\subsection{Related work}
Compute-and-forward can be viewed as one of several wireless instantiations of network coding. Network coding was introduced in \cite{ahlswede:IT00}, where it was shown that network coding achieves the multicast capacity of wireline networks. It was later shown that (random) linear network codes are sufficient for multicast \cite{li:IT03,koetter:TON03,ho:IT06}, and although linear codes are provably insufficient for general wireline networks \cite{dougherty:IT05} they remain popular due to their simplicity and effectiveness. Network coding has been applied to wireless networks by several means. Two information-theoretic techniques are the quantize-map-and-forward of \cite{avestimehr:IT11} and the ``noisy'' network coding of \cite{lim:IT10}, in which relays compress and re-encode the incoming superposition of signals. These approaches generalize the discrete-valued, noiseless combinations of wireline network coding to continuous-valued, noisy combinations over wireless links. For multicast networks, they come to within a constant gap of capacity. Finally, lattice techniques similar to compute-and-forward have been used for the two-way and multi-way relay channels, again achieving rates within a constant gap of capacity \cite{narayanan:Allerton07,nam:IT10,gunduz:IT10,ong:ISIT10}.

Lattice codes play a fundamental role in compute-and-forward. Early works on lattice codes \cite{debuda:JSAC89,linder:IT93,urbanke:IT98} showed that they are sufficient to achieve capacity for the point-to-point AWGN channel. The performance of lattice codes under {\em lattice decoding}---in which the receiver quantizes the incoming signal to the nearest lattice point---was studied in \cite{loeliger:IT97}, and it was shown in \cite{erez:IT04} that lattice decoding achieves capacity. In addition to compute-and-forward, lattice codes have seen use in a variety of information-theoretic problems, including source coding \cite{zamir:IT02,krithivasan:IT09,wagner:IT11}, physical-layer security \cite{he:IT09,agrawal:ISIT09,belifore:ITA10}, and relay networks \cite{nam:IT09,ozgur:ISIT10,nokleby:ICC11,song:IT11}.

Finally, our approach relies heavily on the field of user cooperation. Cooperation was first introduced with the relay channel in \cite{vandermeulen:APP71}. In \cite{cover:IT79} the relay channel is given a thorough treatment, and the most popular relaying strategies---now known as decode-and-forward and compress-and-forward---are presented. More recent work has focused on the diversity gains of cooperation \cite{sendonaris:TOC03a,sendonaris:TON03b,laneman:IT03,laneman:IT04,yuksel:IT07,li:IT11}, showing that cooperating transmitters can obtain diversity gains similar to that of multiple-antenna systems.

\subsection{Notation}
We use bold uppercase letters (e.g. $\mathbf{A}$) to refer to matrices and bold lowercase letters (e.g. $\xbf$) to refer to column vectors. For $n \times m$ matrix $\mathbf{A}$, $\mathbf{a}_i$ refers to the $i$th column of $\mathbf{A}$, i.e. $\mathbf{A} = [\mathbf{a}_1 \cdots \mathbf{a}_m]$. We denote subvectors of a vector using $\xbf[a:b] = (x_a, x_{a+1}, \cdots, x_b)^T$, where $(\cdot)^T$ denotes the usual transpose. We use $\norm{\cdot}$ for the Euclidean norm. Let $\circ$ denote the element-wise or Hadamard product.  Let $\mathbb{F}_p$ denote the finite field of prime characteristic $p$, and let $\oplus$ and $\odot$ denote addition and (matrix) multiplication, respectively, modulo $p$; however, we will occasionally treat the {\em result} of modular arithmetic as a member of the reals according to context. Let $[x]^+ = \max\{x,0\}$ denote the positive part of $x$. Finally, let
\begin{align*}
	C_{\mathrm{mac}}(\mathbf{h},P,\sigma^2) = \min_{\mathcal{B} \subset \{1,\cdots,I\}} \frac{1}{2|\mathcal{B}|}\log\left(1+\frac{P\sum_{i \in \mathcal{B}} h_i^2}{\sigma^2}\right)
\end{align*}
denote the symmetric-rate capacity of the $I$-user Gaussian multiple-access channel having channel gains $\mathbf{h}$ and noise variance $\sigma^2$.

\subsection{Organization}
In Section \ref{sect:prelim} we present the system model and define the performance metrics used in this paper. In Section \ref{sect:main.results} we formally state our main results and provide intuition about their benefits. In Section \ref{sect:lattices} we introduce lattice codes and present the lattice subspace decomposition used in our block Markov strategy. In Section \ref{sect:achievable.rate} we present our cooperative computation strategies in detail and prove that they achieve the computation rates claimed in Section \ref{sect:main.results}. In Section \ref{sect:DMT} we perform a high-SNR analysis of our strategies and prove that they provide the diversity-multiplexing gains claimed in Section \ref{sect:main.results}. In Section \ref{sect:simluations} we present a few numerical examples to showcase the benefits of our approach. Finally, we conclude with Section \ref{sect:conclusion}.

\section{Preliminaries}\label{sect:prelim}
\subsection{System model}
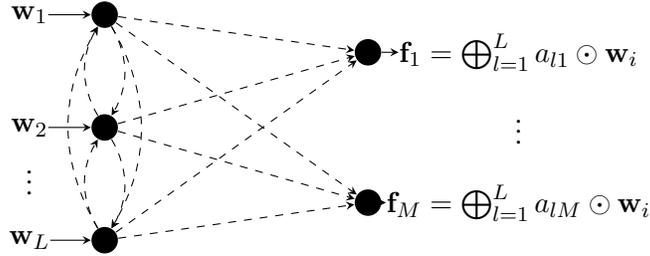
\begin{figure}[htb]
\centering
\begin{tikzpicture}
		[scale=0.5,>=stealth,every node/.style={inner sep=0, minimum size=10}]
		\node (w1) at (-4,3) {$\wbf_1$};
		\node[circle,fill] (x1) at (-2,3) {};
		\node (w2) at (-4,0) {$\wbf_2$};
		\node[circle,fill] (x2) at (-2,0) {}; 
		\node at (-4,-1.25) {$\vdots$};
		\node (wN) at (-4,-3) {$\wbf_L$};
		\node[circle,fill] (xN) at (-2,-3) {};
				
		\draw [->] (w1) -- (x1);
		\draw [->] (w2) -- (x2);
		\draw [->] (wN) -- (xN);
		
		\draw [dashed,->] (x1) to [bend left] (x2);
		\draw [dashed,->] (x1) to [bend left] (xN);
		\draw [dashed,->] (x2) to [bend left] (x1);
		\draw [dashed,->] (x2) to [bend left] (xN);
		\draw [dashed,->] (xN) to [bend left] (x1);
		\draw [dashed,->] (xN) to [bend left] (x2);

		\node[circle,fill] (y1) at (5,2) {};
		\node[circle,fill] (y2) at (5,-2) {};
		\node at (9,0.1) {$\vdots$};
		
		\node (f1) at (9,2) {$\fbf_1=\bigoplus_{l=1}^L a_{l1} \odot \wbf_i$};
		\node (f2) at (9,-2) {$\fbf_M=\bigoplus_{l=1}^L a_{lM} \odot \wbf_i$};
		
		\draw [->] (y1) -- (f1);
		\draw [->] (y2) -- (f2);
		
		\draw [dashed,->] (x1) to (y1);
		\draw [dashed,->] (x2) to (y1);
		\draw [dashed,->] (xN) to (y1);
		
		\draw [dashed,->] (x1) to (y2);
		\draw [dashed,->] (x2) to (y2);
		\draw [dashed,->] (xN) to (y2);
		

\end{tikzpicture}
	\caption{The cooperative compute-and-forward network. $L$ users cooperatively transmit to $M$ receivers, which decode the desired linear functions. \label{fig:system.model}}
\end{figure}

In the {\em cooperative compute-and-forward network}, depicted in Figure \ref{fig:system.model}, $L$ transmitters communicate with $M \leq L$ receivers over the wireless medium. Each of the $L$ users has $T$ messages
$\wbf_l(t) \in \mathbb{F}_p^k$, for $1 \leq t \leq T$. Structurally, this network resembles the compound multiple-access channel or, when $M=L$, the interference channel. However, unlike those more traditional networks, here each receiver intends to decode a finite-field linear combination\footnote{Very precisely, receivers compute any of a {\em sequence} of linear combinations since, as we shall see, $k,p \to \infty$ as the codeword length becomes large.} of the transmitters' messages:
\begin{equation}
	\fbf_m(t) = \bigoplus_{l=1}^L a_{lm} \odot \wbf_l(t),
\end{equation}
for $a_{lm} \in \mathbb{Z}$. Let the matrix $\mathbf{A} = [a_{lm}] \in \mathbb{Z}^{L \times M}$ describe the functions computed by the receivers.

We divide transmissions into $T+1$ blocks of $n$ channel uses each. At block $t$, each transmitter $l$ broadcasts a signal $\xbf_l(t)~\in~\mathbb{R}^n$, subject to an average power constraint:
\begin{equation*}
	\frac{1}{n}\norm{\xbf_l(t)}^2 \leq P,
\end{equation*}
for some $P > 0$. 
The superposition of the transmitters' signals, scaled by channel coefficients and corrupted by noise, arrives at each receiver:
\begin{equation}
	\ybf_m(t) = \sum_{l=1}^Lh_{lm}\xbf_l(t) + \nbf(t),
\end{equation}
where $h_{lm} \in \mathbb{R}$ is the channel coefficient from transmitter $l$ to receiver $m$, and $\nbf(t)$ is a white, unit-variance Gaussian random vector. For convenience, we gather the channel coefficients into the matrix $\mathbf{H} = [h_{lm}]$.

Each transmitter $l$ also obtains the noisy superposition of the other transmitters' signals:
\begin{equation}
	\zbf_l(t) = \sum_{\substack{l^\prime=1 \\ l^\prime \neq l}}^Lg_{l^\prime l}\xbf_{l^\prime}(t)+\nbf_l(t),
\end{equation}
where $g_{l^\prime l} \in \mathbb{R}$ is the channel coefficient from transmitter $l^\prime$ to transmitter $l$, and $\nbf_l(t)$ is again white, unit-variance Gaussian. Again we stack the channel coefficients into a matrix $\mathbf{G} = [g_{l^\prime l}]$ with diagonal elements equal to zero. The choice of zero for the diagonal elements implies  {\em full-duplex} operation, meaning that transmitters can transmit and receive simultaneously. We further assume that channel matrices $\mathbf{H}$ and $\mathbf{G}$ are fixed and known globally among the transmitters and receivers.

We will refer occasionally to the {\em non-cooperative compute-and-forward network}, which is identical to the cooperative network except that the transmitters have no access to each other's transmissions. We model this formally by fixing $\mathbf{G}$ as the all-zero matrix.

\subsection{Computation capacity}
We are interested in the {\em computation capacity} of the network. Since the receivers recover functions of incoming messages, rather than the messages themselves, the computation capacity is defined somewhat differently than the capacity of ordinary channels. We endow each transmitter with an encoder $E_l: \mathbb{F}_p^{k \times T} \times \mathbb{R}^{n \times T} \to \mathbb{R}^{n \times (T+1)}$. That is, the encoder $E_l$ takes as its input the messages $\mathbf{w}_l(t)$ and the received signals $\mathbf{z}_l(t)$ and generates as its output the codewords $\mathbf{x}_l(t)$. We impose a causality restriction on $E_l$: the output codeword $\mathbf{x}_l(t)$ may depend on received signals $\mathbf{z}_l(s)$ only for $s < t$. As usual, the encoding rate is defined as the logarithm of the cardinality of the message set divided by the number of channel realizations over which the messages are encoded:
\begin{equation}
	R = \frac{T \log_2(|\mathbb{F}_p^k|)}{n(T+1)} = \frac{T k \log_2(p)}{n(T+1)} \approx \frac{k \log_2(p)}{n},
\end{equation}
where the approximation holds for large $T$. Note that this is the {\em symmetric} rate among all transmitters.

We endow each receiver with a decoder $D_m : \mathbb{R}^{n \times (T+1)} \to \mathbb{F}_p^{k \times T}$, taking as inputs the received signals $\mathbf{y}_m(t)$ and generating as outputs the estimates $\hat{\mathbf{f}}_m(t)$. Let the absolute probability of error be the probability that any receiver makes an incorrect estimate of any of the desired functions:
\begin{equation}
	P_e = \mathrm{Pr}\{ \hat{\mathbf{f}}_m(t) \neq \mathbf{f}_m(t), \text{for any } 1 \leq m \leq M, 1 \leq t \leq T \}.
\end{equation}
We say that a computation rate $R$ is {\em achievable} if for any $\epsilon > 0$ there exists a sequence of encoders with encoding rate greater than $R - \epsilon$ and decoders such that $P_e \to 0$ as $n \to \infty$. For fixed channel gains $\mathbf{H},\mathbf{G}$, function coefficients $\mathbf{A}$, and transmit power $P$, let $R(\mathbf{H},\mathbf{G},\mathbf{A},P)$ denote the supremum over all achievable computation rates.

In order to define the computation capacity we need to place limitations on the permissible function coefficients $\mathbf{A}$. Otherwise we could choose a trivial coefficient matrix, such as the all-zero matrix, for which the achievable computation rate is unbounded. We therefore require that $\mathbf{A}$ be a member of the following set:
\begin{equation}
	\mathcal{A} = \{ \mathbf{A} \in \mathbb{Z}^{L \times M} : \mathrm{rank}(\mathbf{A}) = M, \ \forall \ m \ \exists \ l \text{ such that } a_{ml} \neq 0\}.
\end{equation}
The first condition ensures that the recovered functions retain as much information as possible about the individual transmitters' messages; for $L=M$ it implies that one can recover the individual messages from the recovered functions. The second condition, which is redundant for $L=M$, ensures that each transmitter is represented in the recovered messages; the receivers cannot simply ignore a transmitter in order to achieve a higher computation rate.

Finally, we define the {\em computation capacity} as the supremum of achievable rates over the set of permissible coefficient matrices:
\begin{equation}\label{eqn:computation.capacity}
	C(\mathbf{H},\mathbf{G},P) = \sup_{\mathbf{A} \in \mathcal{A}} R(\mathbf{H},\mathbf{G},\mathbf{A},P).
\end{equation}

In their seminal work, Nazer and Gastpar developed a computation strategy based on nested lattice codes \cite{nazer:IT11}. It achieves the following computation rate:
\begin{equation}\label{eqn:non.cooperative.rate}
	R_{\mathrm{nc}}(\mathbf{H},P) = \max_{\mathbf{A} \in \mathcal{A}} \min_{1 \leq m \leq M} \left[ \frac{1}{2} \log_2(1 + P\norm{\mathbf{h}_m}^2) - \frac{1}{2}\log_2(\norm{\mathbf{a}_m}^2 + P(\norm{\mathbf{a}_m}^2\norm{\mathbf{h}_m}^2 - |\mathbf{a}_m^T\mathbf{h}_m|^2))\right]^+.
\end{equation}
The first term in (\ref{eqn:non.cooperative.rate}) corresponds to the power in the received signal, whereas the second term is a penalty determined by the gap in the Cauchy-Schwarz inequality between $\mathbf{h}_m$ and $\mathbf{a}_m$. The closer $\mathbf{h}_m$ and $\mathbf{a}_m$ are to being co-linear, the smaller is the rate penalty. Since the Nazer-Gastpar scheme was designed for a non-cooperative network, the rate does not depend on $\mathbf{G}$; nevertheless, it serves as a lower bound on the cooperative computation capacity $C(\mathbf{H},\mathbf{G})$. In the sequel we present a cooperative computation strategy based in part on the Nazer-Gastpar scheme.

\subsection{Diversity-multiplexing tradeoff}
One advantage of user cooperation is that cooperating transmitters can achieve performance similar to that of a multiple-antenna transmitter. Multiple antennas can improve performance on two fronts: increased reliability in the presence of slow channel fading, and increased throughput. In the high-SNR regime, the {\em diversity-multiplexing tradeoff} quantifies this improvement \cite{zheng:IT03}. Let the elements of $\mathbf{H}$ and $\mathbf{G}$ be identically and independently distributed according to a Rayleigh distribution. Next, suppose there is a scheme that achieves the computation rate $R_{\mathrm {scheme}}(\mathbf{H},\mathbf{G},P)$. Then, the {\em diversity order at multiplexing gain $r$} is defined as
\begin{equation}\label{eqn:dmt}
	d(r) = \lim_{P \to \infty} \frac{\log \mathrm{Pr}\{R_{\mathrm{scheme}}(\mathbf{H},\mathbf{G},P) < \frac{r}{2} \log(P)\}}{\log P}.
\end{equation}
In other words, $d(r)$ is the exponent of the outage probability, with the rate taken to have multiplexing gain $r$, as the SNR goes to infinity. The diversity-multiplexing tradeoff of the system, denoted by $d^*(r)$, is the supremum of $d(r)$ over all possible schemes.

The multiplexing gain for compute-and-forward is studied in \cite{niesen:ISIT11}. There it is shown that, using the Nazer-Gastpar approach, the multiplexing gain can be no higher than $\frac{2}{L+1}$. In other words,  $d(r) = 0$ for $r > \frac{2}{L+1}$ for this scheme. In the sequel we show that we can achieve a better diversity-multiplexing tradeoff, including a multiplexing gain of unity, for both cooperative and non-cooperative networks having a single receiver.

\section{Main Results}\label{sect:main.results}
\subsection{Upper bounds}\label{sect:upper.bounds}
First we present two upper bounds on the cooperative computation rate, the proofs of which are contained in the Appendix \ref{app:upper.bounds}. We obtain the first bound by supposing that the transmitters are capable of perfect cooperation, which is equivalent to having a genie supply all messages to each transmitter. The problem then reduces to a multiple-input, single-output (MISO) broadcast channel, the capacity of which is known \cite{weingarten:IT06}. In the sequel we use this result to bound the diversity-multiplexing tradeoff.
\begin{theorem}\label{thm:miso.upper.bound}
	Let the capacity region of a Gaussian MISO broadcast channel be denoted by
	\begin{align}\label{eqn:dirty.paper}
		\mathcal{C}_{\mathrm{miso}}(\mathbf{H},P) = \mathrm{conv} \left\{ \bigcup_{\pi \in \Pi} \left\{\mathbf{r} : r_m \leq \frac{1}{2} \log_2\left(1 +  \frac{\mathbf{h}_{\pi(m)}\mathbf{V}_{\pi(m)}\mathbf{h}_{\pi(m)}^T}{\mathbf{h}_{\pi(m)}\sum_{i=1}^{m-1}\mathbf{V}_{\pi(i)}\mathbf{h}_{\pi(m)}^T+1}\right)\right\}\right\},
	\end{align}
	where $\mathrm{conv}\{\cdot\}$ is the convex hull, $\Pi$ is the set of permutations from $\{1,\cdots L\}$ to itself, and $\mathbf{V}_m$ is a collection of positive semi-definite matrices such that $\sum_{m=1}^M\mathrm{tr}(\mathbf{V}_m) \leq NP$. Then the computation capacity of the cooperative compute-and-forward network is bounded above by
	\begin{equation}
		C(\mathbf{H},\mathbf{G},P) \leq R^+_{\mathrm{miso}}(\mathbf{H},P),
	\end{equation}
	where
	\begin{align}\label{eqn:common.miso.rate}
		R^+_{\mathrm{miso}}(\mathbf{H},P) = \sup \{r : r\mathbf{1} \in \mathcal{C}_{\mathrm{miso}}(\mathbf{H},P) \}
	\end{align}
	is the symmetric-rate capacity of the Gaussian MISO broadcast channel.
\end{theorem}

We obtain the next bound by supposing a genie supplies to the receivers all messages except for those of a single transmitter $l$. Then the receivers need only to recover the messages of transmitter $l$ in order to compute any suitable set of functions. This converts the system to a compound relay channel in which the other transmitters serve as dedicated relays; we bound the capacity of this channel using cut-set arguments. This upper bound is somewhat more realistic than $R^+_{\mathrm{miso}}$, and we use it in Section \ref{sect:simluations} for comparisons to our achievable rates.
\begin{theorem}\label{thm:relay.upper.bound}
	For each transmitter $1 \leq l \leq L$, let $\mathbf{S}_{l} = \{1,\cdots,l-1,l+1,\cdots,l\}$ be the set of transmitters {\em other} than transmitter $l$. Then the computation capacity of the cooperative compute-and-forward network is bounded above by
	\begin{equation*}
		C(\mathbf{H},\mathbf{G},P) \leq R_{\mathrm{single}}^+(\mathbf{H},\mathbf{G},P),
	\end{equation*}
	where
	\begin{equation}
		R_{\mathrm{single}}^+(\mathbf{H},\mathbf{G},P) = \min_{1 \leq l \leq L} \max_{\mathbf{A} \in \mathcal{A}}  \min_{m, a_{lm} \neq 0} \max_{p(\mathbf{x})} \min_{S \in \mathcal{S}_l} I(x_l,x_{S}; y_m,z_{S^C} | x_{S^C}),
	\end{equation}
	where $p(\mathbf{x})$ is any distribution over the transmitted signals $(x_1, \cdots x_L)^T$ satisfying the input power constraint. 
\end{theorem}

\subsection{Achievable rates}
Here we present the computation rates achieved by our cooperative strategy. Our approach is decode-and-forward in nature: at one block transmitters send out lattice codewords corresponding to their individual messages; these messages are decoded by other transmitters. At the next block transmitters cooperatively encode resolution information to assist the receivers. As with any decode-and-forward strategy, we must contend with the fact that it may be difficult for transmitters to decode each other's messages. We therefore require only some of the transmitters to cooperate.\footnote{Other approaches are possible. For example, in an earlier work \cite{nokleby:ITW11} we partitioned the set of transmitters into clusters; transmitters would decode only in-cluster messages. In the interests of brevity we discuss only the approach presented in Theorem \ref{thm:subset}.} A subset $\mathcal{B}$ of the transmitters decodes the messages of every other user, after which they cooperatively transmit resolution information to the receivers. Transmitters not in $\mathcal{B}$, not having decoded incoming messages, do not send any resolution information. We present the details of this strategy, as well as an achievability proof, in Section \ref{sect:achievable.rate}.
 

\begin{theorem}\label{thm:subset}
	Let $\mathcal{B} \subset \{1, \cdots, L\}$. In the cooperative compute-and-forward network, the following computation rate is achievable:
	\begin{multline}\label{eqn:subset}
		R_{c}(\mathbf{H},\mathbf{G},P) = \max_{\mathbf{A} \in \mathcal{A}} \min\bigg\{ \min_{l \in \mathcal{B}} C_{\mathrm{mac}}(\mathbf{g}_l[1:l-1,l+1:L] \circ \mathbf{v}_0[1:l-1,l+1:L], P, 1), \\
		\min_{1 \leq m \leq M} \bigg\{ \frac{1}{2} \log_2\left(1+ \frac{P|\hbf_m^T \vbf_m|^2}{1 + I_{m,r}}\right) + \\
		\left [\frac{1}{2} \log_2(\norm{ P \hbf_m \circ \vbf_0}^2 + I_{m,v}) -
		\frac{1}{2}\log_2\left(\norm{\mathbf{a}_m}^2(1+I_{m,v}) + P\left(\norm{\mathbf{a}_m}^2\norm{\mathbf{h}_m \circ \mathbf{v}_0}^2 - |\mathbf{a}_m^T(\mathbf{h}_m \circ \mathbf{v}_0)|^2\right)\right)\right]^+ \bigg\} \bigg\},
	\end{multline}
	where
	\begin{equation}
		I_{m,r} = P\left(\norm{ \hbf_m \circ \vbf_0 }^2 + \sum_{m^\prime \neq m,0} |\hbf^T_m \vbf_{m^\prime} |^2\right)
	\end{equation}
	is the interference power seen at receiver $m$ as it decodes its resolution information,
	\begin{equation}
		I_{m,v} =  P\sum_{m^\prime \neq m,0} |\hbf_m^T \vbf_{m^\prime}|^2
	\end{equation}
	is the interference seen at receiver $m$ as it decodes the vestigial information, and for any vectors $\vbf_0,\vbf_1,\cdots, \vbf_M$ such that
	\begin{equation}
		\sum_{m=0}^M |v_{lm}|^2 \leq 1, \forall l
	\end{equation}
	and $v_{lm} = 0$ for  $l \neq \mathcal{B}$ and $m>0$.
\end{theorem}

The achievable rate (\ref{eqn:subset}) is a bit difficult to parse, so we take time here to describe each of its three components. First is the rate of a Gaussian multiple-access channel, which corresponds to the rate at which cooperating transmitters can decode others' messages. Second is the rate at which each receiver can decode the resolution information, which is that of a virtual MISO link between cooperating transmitters and the receiver; signals unrelated to the resolution information are treated as noise. Third is the rate at which the receivers, having already decoded the resolution information, can decode the vestigial component of the desired combination of lattice points; this is  the Nazer-Gastpar rate of (\ref{eqn:non.cooperative.rate}), with resolution information intended for other receivers treated as noise.

Each transmitter splits its power between sending its own lattice codewords and cooperatively sending resolution information. The split is defined by the steering vectors $\mathbf{v}_0, \mathbf{v}_1, \cdots, \mathbf{v}_M$. Each element $v_{l0}$ dictates the fraction of power transmitter $l$ expends on its own lattice codewords. For cooperating transmitter $l$, each element $v_{lm}$ dictates the fraction of power expended on resolution information for receiver $m$. The steering vectors introduce two separate notions of alignment. First, we can choose $\mathbf{v}_0$ in order to minimize the Cauchy-Schwarz penalty in (\ref{eqn:subset}). Second, we can choose the remaining vectors $\mathbf{v}_m$ to trade off between increasing the coherence gain at the intended receivers and decreasing the interference generated at other receivers. Finding the optimum steering vectors is a non-covex problem; for further results and in our simulations we rely on a few heuristic means for selecting them.

We can obtain a simpler expression for the achievable rate by choosing $\mathcal{B} = \{1, \cdots, L\}$ and taking the steering vectors $\mathbf{v}_1, \cdots, \mathbf{v}_M$ to be zero-forcing beamformers. Thus the cooperative signals do not interfere at other receivers.
\begin{corollary}\label{cor:zero.forcing}
	The following computation rate is achievable for the cooperative compute-and-forward network:
	\begin{multline}
		R_{\mathbf{zf}}(\mathbf{H},\mathbf{G},P) = \max_{\mathbf{A} \in \mathcal{A}} \min\bigg\{ \min_{1 \leq l \leq L}C_{\mathrm{mac}}(\mathbf{g}_l[1:l-1,l+1:L] \circ \mathbf{v}_0[1:l-1,l+1:L], P, 1),\\
			\min_{1 \leq m \leq M}  \bigg[\frac{1}{2}\log_2(1+P(\norm{\mathbf{h}_m\circ \mathbf{v}_0}^2 + |\mathbf{h}_m^T\mathbf{v}_m|^2)) - \\
			\frac{1}{2}\log_2\left(\norm{\mathbf{a}_m}^2 + P\left(\norm{\mathbf{a}_m}^2\norm{\mathbf{h}_m \circ \mathbf{v}_0}^2 - |\mathbf{a}_m^T(\mathbf{h}_m \circ \mathbf{v}_0)|^2\right)\right)\bigg]^+ \bigg\},
	\end{multline}
	for any vectors $\mathbf{v}_0, \mathbf{v}_1, \cdots, \mathbf{v}_M$ satisfying
	\begin{equation}
		\sum_{m=0}^M |v_{lm}|^2 \leq 1
	\end{equation}
	and
	\begin{equation}
		\mathbf{v}_m^T\mathbf{h}_{m^\prime} = 0, \forall \ m \neq m^\prime.
	\end{equation}
\end{corollary}
Since $L \geq M$, it is possible to choose non-trivial zero-forcing beamforming vectors for almost every $\mathbf{H}$.

Finally, choosing $\mathcal{B} = \emptyset$, we obtain an achievable rate for both the cooperative and non-cooperative compute-and-forward network. This yields a rate similar to (\ref{eqn:non.cooperative.rate}), except that each transmitter can adjust its transmit power in order to tune the effective channels to match the desired linear functions. In fact this rate is a special case of the ``superposition'' compute-and-forward presented in \cite[Theorem 13]{nazer:IT11}.
\begin{corollary}\label{cor:non.cooperative}
	In both the non-cooperative compute-and-forward network and the cooperative compute-and-forward network, the following rate is achievable:
	\begin{multline}
		R(\mathbf{H},\mathbf{G},P) = \max_{\mathbf{A} \in \mathcal{A}} \min_{1 \leq m \leq M} \bigg[\frac{1}{2}\log_2(1+P(\norm{\mathbf{h}_m\circ \mathbf{v}_0}^2)) - \\ \frac{1}{2}\log_2\left(\norm{\mathbf{a}_m}^2 + P\left(\norm{\mathbf{a}_m}^2\norm{\mathbf{h}_m \circ \mathbf{v}_0}^2 - |\mathbf{a}_m^T(\mathbf{h}_m \circ \mathbf{v}_0)|^2\right)\right)\bigg]^+,
	\end{multline}
	for any $\mathbf{v}_0$ satisfying
	\begin{equation}
		|v_{l0}|^2 \leq 1, \forall \ 1 \leq l \leq L.
	\end{equation}
\end{corollary}

\subsection{Diversity-multiplexing tradeoff}
Here we present our diversity-multiplexing tradeoff results, the proofs of which are presented in Section \ref{sect:DMT}. We begin with the non-cooperative case.
\begin{theorem}\label{thm:non.cooperative.dmt}
	For the non-cooperative compute-and-forward network, the diversity-multiplexing tradeoff for any scheme is upper-bounded as follows:
	\begin{equation}
		d^*(r) \leq d^+_{\mathrm{nc}}(r) = 1-r.
	\end{equation}
	For the case of $M=1$, $d^*(r) = d^+_{\mathrm{nc}}(r)$.
\end{theorem}
In other words, the DMT of the non-cooperative compute-and-forward network is bounded above by that of a scalar Gaussian channel. In the case of a single receiver, we can achieve this upper bound with lattice codes and signal alignment. With the steering vector $\mathbf{v}_0$ chosen such that the equivalent channel vector is a constant, the achievable rate---and therefore the error probability---is approximately that of a single SISO link.

Next, we look at the DMT of the cooperative compute-and-forward network. We start by presenting an upper bound.
\begin{theorem}\label{thm:cooperative.dmt}
	For the cooperative compute-and-forward network, the diversity-multiplexing tradeoff is upper-bounded as
	\begin{equation}
		d^*(r) \leq d^+_\mathrm{c}(r) = L(1-r). 
	\end{equation}
\end{theorem}
In other words, the DMT is upper-bounded by that of a single MISO link. In the case of a single receiver, we derive two lower bounds on the DMT. The first is derived using a rather simple strategy employing time sharing and Gaussian codes.
\begin{theorem}\label{thm:random.dmt}
	For the cooperative compute-and-forward network, the following diversity-multiplexing tradeoff is achievable:
	\begin{equation}
		d_{\mathrm{random}}^-(r) = L \min\{1-2r, (L-1)(1 - 2(L-1)r) \}.
	\end{equation}
	In particular, $d_{\mathrm{random}}^-(0) = L$.
\end{theorem}
Since it involves time sharing, the strategy used in Theorem \ref{thm:random.dmt} has poor multiplexing performance. It does, however, achieve the full diversity gain of $L$. The second bound is derived using the cooperative computation strategy of Theorem \ref{thm:subset}.

\begin{theorem}\label{thm:lattice.dmt}
	For the cooperative compute-and-forward network, the following diversity-multiplexing tradeoff is achievable:
	\begin{multline}
		d_{\mathrm{lattice}}^-(r) = 1-r + \min\{[1-2r]^+,[(L-1)(1-rL)]^+ \} + \\  \max_{0 \leq x \leq 1}(L-2) \min\{[1-x-r]^+,[(L-1)(1-(L-1)r-x)]^+,[x-r]^+ \}.
	\end{multline}
	Here, $d_{\mathrm{lattice}}^-(0) = 2 + \frac{L-2}{2}$.
\end{theorem}

Here the main difficulty is the Cauchy-Schwarz penalty inherent to lattice coding. It turns out that choosing $\mathbf{v}_0$ to align with the channels, as we did in the non-cooperative case, precludes cooperation with high probability. We therefore choose $\mathbf{v}_0$ to be constant, taking the Cauchy-Schwarz penalty ``on the chin.'' We balance the transmit power between sending fresh information, which helps transmitters decode others' messages, and sending resolution information, which helps the receiver decode the desired linear combination. Choosing the balance properly, the benefits of cooperation outweigh the Cauchy-Schwarz penalty, but only enough to obtain a diversity gain of approximately $1/2$ per transmitter. Nevertheless, for higher multiplexing gains lattice coding outperforms the strategy of Theorem \ref{thm:random.dmt}.

\begin{figure}[htb]
	\centering
\begin{tikzpicture}

	\begin{axis}[
	title={$L = 2$ Transmitters},
	xlabel={Multiplexing gain $r$},
	ylabel={Diversity gain $d$},
	axis x line=bottom,
	axis y line=left,
	height=200pt,
	width=250pt
	]
		\addplot[smooth,color=red, very thick,solid] file {trans/DMT.2.siso};
		\addlegendentry{\small Non-cooperative lower bound};	
		
		\addplot[,color=black,very thick,densely dotted] file {trans/DMT.2.random};
		\addlegendentry{\small Cooperative random coding};
		
		\addplot[,color=blue,very thick,dashed] file {trans/DMT.2.coop};
		\addlegendentry{\small Cooperative lattice coding};
		
		\addplot[color=green!50!black,very thick,dotted] file {trans/DMT.2.miso};
		\addlegendentry{\small MISO upper bound};
	\end{axis}
	
	\begin{scope}[xshift=275pt]
			\begin{axis}[
	title={$L=5$ Transmitters},
	xlabel={Multiplexing gain $r$},
	ylabel={Diversity gain $d$},
	axis x line=bottom,
	axis y line=left,
	height=200pt,
	width=250pt
	]

		\addplot[smooth,color=red, very thick,solid] file {trans/DMT.5.siso};
		\addlegendentry{\small Non-cooperative lower bound};	

		\addplot[,color=black,very thick,densely dotted] file {trans/DMT.5.random};
		\addlegendentry{\small Cooperative random coding};
		
		\addplot[color=blue,very thick,dashed] file {trans/DMT.5.coop};
		\addlegendentry{\small Cooperative lattice coding};
		
		\addplot[color=green!50!black,very thick,dotted] file {trans/DMT.5.miso};
		\addlegendentry{\small MISO upper bound};
	\end{axis}
	\end{scope}
\end{tikzpicture}
	\caption{Diversity-multiplexing tradeoff for $L=2$, $L=5$ transmitters and a single reciever.}
	\label{fig:dmt.plot}
\end{figure}
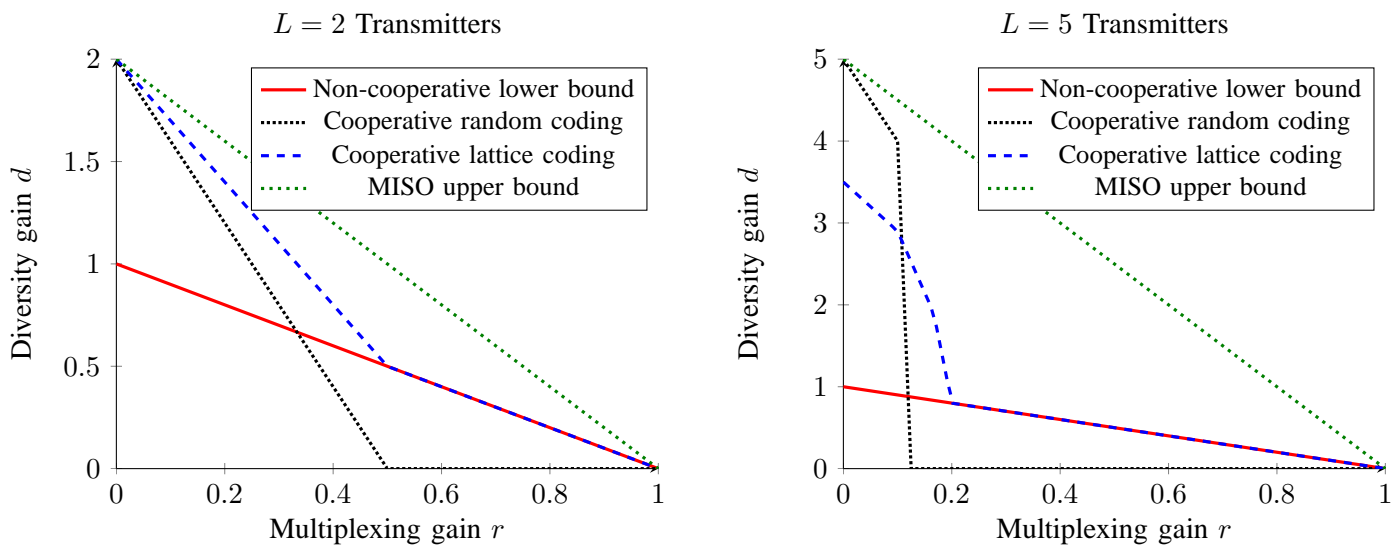

We plot the DMT bounds in Figure \ref{fig:dmt.plot}. For $L=2$ lattice coding is sufficient to achieve full diversity, and the DMT achieved by lattice coding strictly dominates that achieved by random coding. For $L>2$, lattice coding achieves better performance only for sufficiently high multiplexing gain. Random coding fails altogether at multiplexing gains higher than $(L-1)/2$ due to the need for transmitters to decode $L-1$ separate messages and the need for time-sharing. Lattice coding, on the other hand, maintains non-zero diversity for every any $0 \leq r \leq 1$. Between the two strategies we obtain the corner points of the DMT region.

\section{Lattice codes}\label{sect:lattices}
In this section we first introduce the basics of lattice codes, after which we present the lattice decomposition employed in our cooperative computation strategy.

\subsection{Capacity-achieving lattice codes}
Formally, a lattice $\Lambda$ is a discrete additive subgroup of $\mathbb{R}^n$, which implies that for any $\lambda_1, \lambda_2 \in \Lambda$ we have $\lambda_1 + \lambda_2 \in \Lambda$ and $\lambda_1 - \lambda_2 \in \Lambda$. Any lattice can be generated by taking integer combinations of (not necessarily unique) basis vectors. Choosing these basis vectors as columns, we form the {\em generator matrix} of  $\Lambda$, denoted by $\mathbf{G} \in \mathbb{R}^{n\times n}$:
\begin{equation}
	\Lambda = \mathbf{G}\mathbb{Z}^n.
\end{equation}
We let $Q_{\Lambda}$ denote the {\em lattice quantizer}, which maps any point $\xbf \in \mathbb{R}^n$ to the nearest point in $\Lambda$:
\begin{equation}
	Q_{\Lambda}(\mathbf{x}) = \arg\min_{\lambda \in \Lambda} \norm{\mathbf{x} - \lambda}.
\end{equation}
The lattice $\Lambda$ induces a partition of $\mathbb{R}^n$ into the {\em Voronoi regions} $\Vcal(\lambda)$ of each lattice point $\lambda \in \Lambda$:
\begin{equation}
	\Vcal(\lambda) = \{\mathbf{x} \in \Re^n : Q_{\Lambda}(\mathbf{x}) = \lambda \},
\end{equation}
where ties are broken arbitrarily. In other words, the Voronoi region of $\lambda \in \Lambda$ is simply the set of points that are closer to $\lambda$ than to any other lattice point.

Let $\Vcal = \Vcal(0)$ be the {\em fundamental Voronoi region} of $\Lambda$. The$\mod$operation with respect to $\Lambda$ returns the quantization error
\begin{equation}
	\mathbf{x} \mod \Lambda = \mathbf{x} - Q_{\Lambda}(\mathbf{x}),
\end{equation}
which is always a member of $\Vcal$. The$\mod$operation allows one to draw an analogy with modulo arithmetic over a finite field. Just as modulo arithmetic ensures that the result remains a member of the finite field, performing arithmetic modulo $\Lambda$ ``wraps'' the result within $\Vcal$. The$\mod$operation obeys the associativity property:
\begin{equation}
	[[\xbf] \mod \Lambda + \ybf] \mod \Lambda = [\xbf + \ybf] \mod \Lambda.
\end{equation}

The {\em second moment} $\sigma^2(\Lambda)$ quantifies the average power of a random variable uniformly distributed inside $\Vcal$:
\begin{equation}
	\sigma^2(\Lambda) = \frac{1}{n \mathrm{Vol}(\Vcal)}\int_{\Vcal} \norm{\xbf}^2 d\xbf,
\end{equation}
where $\mathrm{Vol}(A)$ is the volume of a set $A \subset \Re^n$.
The {\em normalized second moment} is defined as:
\begin{equation}
	G(\Lambda) = \frac{\sigma^2(\Lambda)}{\mathrm{Vol}(\Vcal)^{\frac{2}{n}}}.
\end{equation}
The normalized second moment provides a measure of the efficiency of $\Lambda$ as a shaping region. The closer $\Vcal$ is to being spherical, the smaller $G(\Lambda)$ is.

The {\em covering radius} $r_{\mathrm{cov}}(\Lambda)$ is the radius of the smallest sphere that covers $\Vcal$:
\begin{equation}
	r_{\mathrm{cov}}(\Lambda) = \inf_r \{r>0 | \Vcal \subset r \mathcal{B}_n \},
\end{equation}
where $\mathcal{B}_n$ is the unit sphere in $\mathbb{R}^n$. The {\em effective radius} $r_{\mathrm{eff}}(\Lambda)$ be the radius of a sphere with the same volume as $\Vcal$:
\begin{equation}
	r_{\mathrm{eff}}(\Lambda) = \left( \frac{\mathrm{Vol}(\Vcal)}{\mathrm{Vol}(\mathcal{B}_n)}\right)^{\frac{1}{n}}.
\end{equation}
Note that $r_{\mathrm{cov}}(\Lambda) \geq r_{\mathrm{eff}}(\Lambda)$.

In order to construct lattice codebooks suitable for proving information-theoretic results, we require {\em sequences} of lattices that asymptotically satisfy several desirable properties. For example, we say that a sequence of lattices $\{\Lambda^{(n)}\}, \Lambda^{(n)} \in \mathbb{R}^n$, is {\em good for covering} or {\em Rogers good} \cite{rogers:M59} provided the covering radius approaches the effective radius:
\begin{equation*}
	\lim_{n \to \infty} \frac{r_{\mathrm{cov}}(\Lambda^{(n)})}{r_{\mathrm{eff}}(\Lambda^{(n)})} = 1.
\end{equation*}
Similarly, a sequence of lattices is {\em good for quantization} provided
\begin{equation*}
	\lim_{n \to \infty} G(\Lambda^{(n)}) = \frac{1}{2\pi e}.
\end{equation*}
Finally, let $\zbf \sim \mathcal{N}(0,\sigma^2\mathbf{I})$ be a Gaussian random vector. Define the {\em volume-to-noise ratio} $\mu(\Lambda,P_e)$ as
\begin{equation*}
	\mu(\Lambda,P_e) = \frac{(\mathrm{Vol}(\Vcal))^{\frac{2}{n}}}{\sigma^2},
\end{equation*}
where $\sigma^2$ is chosen such that $\mathrm{Pr}\{\zbf \notin \Vcal \} = P_e$.
A sequence of lattices $\Lambda^{(n)}$ is {\em good for AWGN coding} or {\em Poltyrev good} if
\begin{equation*}
	\lim_{n \to \infty} \mu(\Lambda^{(n)},P_e) = 2\pi e.
\end{equation*}
The existence of such sequences was proven by Poltyrev in \cite{poltyrev:IT94}. Furthermore, Erez et al. proved that there exist sequences of lattices that are simultaneously good for covering, quantization, and AWGN coding \cite{erez:IT05}.

\begin{figure}[htb]
\centering
\begin{tikzpicture}
		[scale=0.5,
		fine/.style = {solid,draw=gray},
		shaping/.style = {very thick},
		voronoi/.style = {fill=black!10!white}]
		\clip (-5.5,-4.75) rectangle (5.5,4.75); 
		
		\fill[voronoi] (0:3*1.1547) -- (60:3*1.1547) -- (120:3*1.1547) -- (180:3*1.1547) -- (240:3*1.1547) -- (300:3*1.1547) -- (360:3*1.1547);
		
		\foreach \t in {-3,...,3} 
			\foreach \u in {-3,...,3} 
			{ 
				\begin{scope}[xshift = 1.732*\u cm, yshift = 2*\t cm+\u cm]
					\draw[fine]  (0:1.1547) -- (60:1.1547) -- (120:1.1547) -- (180:1.1547) -- (240:1.1547) -- (300:1.1547) -- (360:1.1547);
					\draw (0,0) circle (0.1cm);
				\end{scope}
			} 
			
		\foreach \t in {-1,0,1} 
			\foreach \u in {-1,0,1} 
			{ 
				\begin{scope}[xshift = 3*1.732*\u cm, yshift = 6*\t cm+3*\u cm]
					\draw[shaping]  (0:3*1.1547) -- (60:3*1.1547) -- (120:3*1.1547) -- (180:3*1.1547) -- (240:3*1.1547) -- (300:3*1.1547) -- (360:3*1.1547);
					\fill (0,0) circle (0.1 cm); 
				\end{scope}
			}
		\node at (300:0.65) {$\Vcal_c$};
		\node at (300:2.9) {$\Vcal_s$};
\end{tikzpicture}
	\caption{Nested lattice codes. White dots are elements of the coding lattice, and black dots are elements of the shaping lattice. Each lattice point inside the shaded Voronoi region $\Vcal_s$ is a member of the codebook. \label{fig:lattice}}
\end{figure}
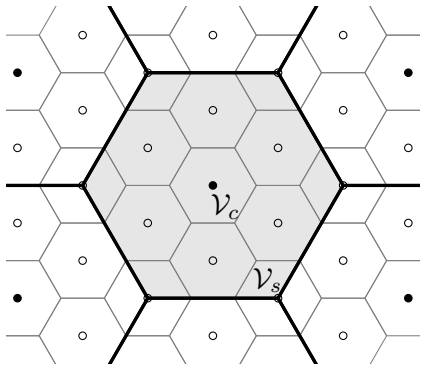

Lattice codebooks are constructed using {\em nested lattices}, as depicted in Figure \ref{fig:lattice}. Here we review the construction of codebooks sufficient to achieve capacity for the AWGN point-to-point channel, which is the model for  codebooks to be used throughout this paper. Let $\Lambda_s^{(n)}$ be a sequence of {\em shaping lattices} that are good for covering and AWGN coding and satisfy $\sigma^2(\Lambda_s^{(n)}) = 1$, and let $\mathbf{G}_s^{(n)}$ denote generator matrices for each lattice in the sequence. Then, following \cite{krithvasan:TR}, we adapt Construction A \cite{loeliger:IT97} to construct a sequence of coding lattices $\Lambda_c^{(n)} \supset \Lambda_s^{(n)}$. The construction process goes as follows:
\begin{enumerate}
	\item For each $n$, choose an integer $k$ and a prime $p$. Draw a $n \times k$ matrix $\mathbf{F}_c^{(n)} \in \mathbb{F}_p^{n \times k}$ randomly and uniformly.
	\item Construct the linear codebook over $\mathbb{F}_p$ defined by $\mathbf{F}_c^{(n)}$:
	\begin{equation*}
		\hat{\mathcal{C}}^{(n)} = \mathbf{F}_c^{(n)}\mathbb{F}_p^k
	\end{equation*}
	\item ``Lift'' the codebook $\hat{\mathcal{C}}^{(n)}$ to $\mathbb{R}^n$ by defining the lattice
	\begin{equation*}
		 \hat{\Lambda}_c^{(n)} = p^{-1}\hat{\mathcal{C}}^{(n)} + \mathbb{Z}^n.
	\end{equation*}
	\item Finally, rotate $\hat{\Lambda}_c^{(n)}$ so that it is nested inside $\Lambda_s^{(n)}$:
	\begin{equation*}
		\Lambda_c^{(n)} = \mathbf{G}_s^{(n)} \hat{\Lambda}_c^{(n)}.
	\end{equation*}
\end{enumerate}
We form the lattice codebook by taking the intersection of the coding lattice with the fundamental Voronoi region of the shaping lattice:
\begin{equation*}
	\mathcal{C}^{(n)} = \Lambda_c^{(n)} \cap \Vcal_s^{(n)}.
\end{equation*}
The rate of this codebook is
\begin{equation*}
	R = \frac{1}{n}\log_2 | \mathcal{C}^{(n)} | = \frac{k \log_2(p)}{n}.
\end{equation*}
It is shown in \cite{nazer:IT11} that choosing $p$ such that $n/p \to 0$ as $n \to \infty$ guarantees that the sequence of coding lattices $\Lambda_c^{(n)}$ is good for AWGN coding. For any desired rate $R > 0$, we can construct an appropriate sequence of codebooks by choosing $p = n \log_2(n)$ and $k = \lfloor \frac{nR}{\log_2(p)} \rfloor$.

In essence, the preceding codebook construction allows us to take a linear block code over $\mathbb{F}_p$ and to create a corresponding linear code over Euclidean space. If the underlying linear code achieves capacity, as does the ensemble of random linear codes, so too does the resulting lattice codebook. We can use any linear code in place of the one chosen above; the performance cost is only the gap to capacity of the linear code chosen.

For the lattice compute-and-forward proposed in \cite{nazer:IT11}, an important fact is that there exists a mapping from finite-field messages to lattice codewords that preserves linearity. That is, the mapping sends finite-field linear combinations of messages to integer sums of lattice points modulo the shaping lattice. Formally, this implies that there is an isomorphism between the additive group of field elements and the group of lattice codewords modulo the shaping lattice. We state this result in the following lemma.
\begin{lemma}\label{lem:isomorphism}
There exists an isomorphism $\phi : \mathbb{F}_p^k \to \Ccal^{(n)}$, namely
\begin{equation}
	\phi(\wbf) = [\Gbf_s^{(n)}p^{-1}\Fbf_c^{(n)}\wbf] \mod \Lambda_s^{(n)}.
\end{equation}
\end{lemma}
\begin{IEEEproof}
We need to show that $\phi$ is a bijection and that it respects the group operation; that is, $\phi(\mathbf{w}_1 \oplus \mathbf{w}_2) = [\phi(\mathbf{w}_1) + \phi(\mathbf{w}_2)] \mod \Lambda_s^{(n)}$ for any $\mathbf{w}_1,\mathbf{w}_2 \in \mathbb{F}_p^k$. That $\phi$ is a bijection was shown in \cite[Lemma 5]{nazer:IT11}. To see that $\phi$ respects the group operation, we resort to direct computation:
\begin{align}
	\phi(\mathbf{w}_1 \oplus \mathbf{w}_2) &= [\Gbf_s^{(n)}p^{-1}\Fbf_c^{(n)}(\mathbf{w}_1 \oplus \mathbf{w}_2)] \mod \Lambda_s^{(n)} \\
	&= [\Gbf_s^{(n)}p^{-1}(\Fbf_c^{(n)}(\mathbf{w}_1 + \mathbf{w}_2) + p\mathbf{i})] \mod \Lambda_s^{(n)},
\end{align}
where $\mathbf{i} \in \mathbb{Z}^n$ is a vector of integers corresponding to the discrepancy between real-valued and modulo-$p$ arithmetic. Continuing, we get
\begin{align}
	\phi(\mathbf{w}_1 \oplus \mathbf{w}_2) &= [\Gbf_s^{(n)}p^{-1}\Fbf_c^{(n)}(\mathbf{w}_1 + \mathbf{w}_2) + \Gbf_s^{(n)}\mathbf{i}] \mod \Lambda_s^{(n)} \\
	&= [\Gbf_s^{(n)}p^{-1}\Fbf_c^{(n)}(\mathbf{w}_1 + \mathbf{w}_2)] \mod \Lambda_s^{(n)} \\
	&= [\phi(\mathbf{w}_1) + \phi(\mathbf{w}_2)] \mod \Lambda_s^{(n)}.
\end{align}
where the last equality is due to the fact that $\Gbf_s^{(n)}\mathbf{i} \in \Lambda_s^{(n)}$ and that adding a member of $\Lambda_s^{(n)}$ does not change the result of the arithmetic modulo $\Lambda_s^{(n)}$.
\end{IEEEproof}

\subsection{Lattice subspaces}\label{sect:lattice.subspaces}
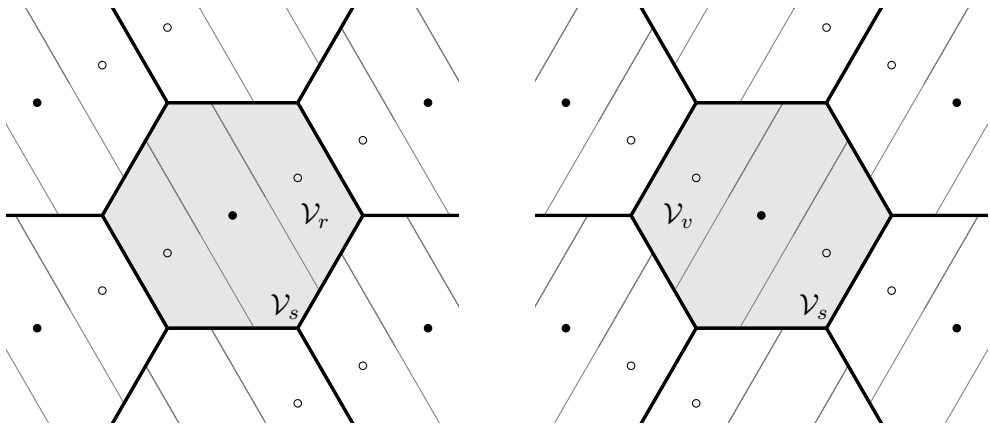
\begin{figure}[htb]
\centering
\begin{tikzpicture}
		[scale=0.5,
		fine/.style = {solid,draw=gray},
		shaping/.style = {very thick},
		voronoi/.style = {fill=black!10!white}]
		\begin{scope}
		\clip (-6,-5.5) rectangle (6,5.5); 
		
		\fill[voronoi] (0:3*1.1547) -- (60:3*1.1547) -- (120:3*1.1547) -- (180:3*1.1547) -- (240:3*1.1547) -- (300:3*1.1547) -- (360:3*1.1547);
			
		\foreach \t in {-1,0,1} 
			\foreach \u in {-1,0,1} 
			{ 
				\begin{scope}[xshift = 3*1.732*\u cm, yshift = 6*\t cm+3*\u cm]					
					\foreach \v in {-1,0,1} 
					{ 
						\begin{scope}[xshift = 1.732*\v cm, yshift = \v cm]
							\draw (0,0) circle (0.1cm);
						\end{scope}
					} 
					
					\foreach \v in {0,1}
					{
						\begin{scope}[xshift = 1.732*\v cm, yshift = \v cm]
							\draw[fine]  (210:1) -- ++(120:2.9) -- ++(120:-5.8);
						\end{scope}
					}
					
					\draw[shaping]  (0:3*1.1547) -- (60:3*1.1547) -- (120:3*1.1547) -- (180:3*1.1547) -- (240:3*1.1547) -- (300:3*1.1547) -- (360:3*1.1547);
					\fill (0,0) circle (0.1 cm); 
					
				\end{scope}
			}
		\node at (2.2,0) {$\Vcal_r$};
		\node at (300:2.8) {$\Vcal_s$};
		\end{scope}
		
		\begin{scope}[xshift = 400]
			\clip (-6,-5.5) rectangle (6,5.5); 
		
			\fill[voronoi] (0:3*1.1547) -- (60:3*1.1547) -- (120:3*1.1547) -- (180:3*1.1547) -- (240:3*1.1547) -- (300:3*1.1547) -- (360:3*1.1547);
			
			\foreach \t in {-1,0,1} 
			\foreach \u in {-1,0,1} 
			{ 
				\begin{scope}[xshift = 3*1.732*\u cm, yshift = 6*\t cm+3*\u cm]
					\foreach \u in {-1,...,1} 
					{ 
						\begin{scope}[xshift = -1.732*\u cm, yshift = \u cm]
							\draw (0,0) circle (0.1cm);
						\end{scope}
					} 
					
					\foreach \u in {-1,0} 
					{ 
						\begin{scope}[xshift = -1.732*\u cm, yshift = \u cm]
							\draw[fine]  (150:1) -- ++(60:2.9) -- ++(60:-5.8);
						\end{scope}
					}

					\draw[shaping]  (0:3*1.1547) -- (60:3*1.1547) -- (120:3*1.1547) -- (180:3*1.1547) -- (240:3*1.1547) -- (300:3*1.1547) -- (360:3*1.1547);
					\fill (0,0) circle (0.1 cm); 
				\end{scope}
			}
		
		\node at (-2.2,0) {$\Vcal_v$};
		\node at (300:2.8) {$\Vcal_s$};
		\end{scope}
\end{tikzpicture}
	\caption{Lattice subspace decomposition. Each lattice codeword in $\Ccal^{(n)}$ is the sum of a point in $\Ccal_r^{(n)}$ (left) and a point in $\Ccal_v^{(n)}$ (right). The shaded region $\Vcal_s$ defines the codebook, whereas the strip-shaped Voronoi regions $\Vcal_r$ and $\Vcal_v$ define the decoding regions of the resolution and vestigial codebooks, respectively. \label{fig:sublattice.code}}
\end{figure}

In our cooperative computation strategy, we tailor block Markov encoding to lattice codes. To do so, we introduce a key ingredient of our approach: the decomposition of the lattice codebook into subspaces. Let $k_r \leq k$, and let $\Fbf_r^{(n)} \in \mathbb{F}_p^{n \times k_r}$ denote the matrix composed of the first $k_r$ columns of $\Fbf_c^{(n)}$. Similarly, let $k_v = k - k_r$, and let $\Fbf_v^{(n)} \in \mathbb{F}_p^{n \times k_v}$ denote the matrix of the remaining $k_v$ columns. Then define the {\em resolution lattice} $\Lambda_r$ and the {\em vestigiall\footnote{This terminology is intended to convey the fact that this lattice component encodes the ``residual'' or ``leftover'' information bits. We use this less-common synonym in order to minimize notational confusion.} lattice} $\Lambda_v$ as
\begin{align*}
	\Lambda_r^{(n)} &= \Gbf_s^{(n)}(p^{-1}\Fbf_r^{(n)}\mathbb{F}_p^{k_r} + \mathbb{Z}^n) \\
	\Lambda_v^{(n)} &= \Gbf_s^{(n)}(p^{-1}\Fbf_v^{(n)}\mathbb{F}_p^{k_v} + \mathbb{Z}^n).
\end{align*}
Since these sequences of lattices are special cases of the lattice construction from the previous subsection, each sequence is individually good for AWGN coding. By construction $\Lambda_c^{(n)} = \Lambda_r^{(n)} + \Lambda_v^{(n)}$ and $\Lambda_s^{(n)} \subset \Lambda_r^{(n)}, \Lambda_v^{(n)} \subset \Lambda_c^{(n)}$. Define the resolution and vestigial codebooks
\begin{align*}
	\Ccal_r^{(n)} &= \Lambda_r^{(n)} \cap \Vcal_{\Lambda_s^{(n)}} \\
	\Ccal_v^{(n)} &= \Lambda^{(n)}_v \cap \Vcal_{\Lambda_s^{(n)}},
\end{align*}
having rates
\begin{align*}
	R_r &= \frac{k_r}{n}\log_2 p \\
	R_v &= \frac{k_v}{n}\log_2 p.
\end{align*}
By construction $R_r + R_v = R_c$. Furthermore, for any $0 \leq R_r \leq R$, we can choose $k_r = \lfloor \frac{nR_r}{\log_2(p)} \rfloor$ to achieve the desired resolution codebook rate. For any message $\mathbf{w} \in \mathbb{F}_p^k$, we can define the {\em projection} onto the resolution and vestigial codebook as follows:
\begin{align*}
	\phi_r(\wbf) &= [\Gbf_s p^{-1}\Fbf_r \wbf[1:k_r]] \mod \Lambda_s \\
	\phi_v(\wbf) &= [\Gbf_s p^{-1}\Fbf_v \wbf[k_r+1:k]] \mod \Lambda_s.
\end{align*}
Using these projections, we can define a linear decomposition of the lattice codebook, as depicted in Figure \ref{fig:sublattice.code}.
\begin{lemma}\label{lem:subspace.decomposition}
For any $\wbf \in \mathbb{F}_p^k$,
\begin{equation}
	\phi(\wbf) = [\phi_r(\wbf) + \phi_v(\wbf)] \mod \Lambda_s^{(n)},
\end{equation}
\end{lemma}
\begin{IEEEproof}
	This result follows from Lemma \ref{lem:isomorphism}. By definition $\wbf = (\wbf^T[1:k_r] \mathbf{0}^T_{k_v})^T \oplus (\mathbf{0}_{k_r}^T \wbf^T[k_r+1:k])^T$, so
	\begin{align*}
		\phi(\wbf) &= \phi((\wbf^T[1:k_r], \mathbf{0}^T_{k_v})^T \oplus (\mathbf{0}_{k_r}^T, \wbf^T[k_r+1:k])^T) \\
		&= [\phi((\wbf^T[1:k_r], \mathbf{0}^T_{k_v})^T) + \phi((\mathbf{0}_{k_r}^T, \wbf^T[k_r+1:k])^T)] \mod \Lambda_s^{(n)} \\
		&= [\phi_r(\wbf) + \phi_v(\wbf)] \mod \Lambda_s^{(n)},
	\end{align*}
	where the last equality follows from the definition of $\Fbf_r^{(n)}$ and $\Fbf_v^{(n)}$; zeroing out the unwanted portions of $\wbf$ is equivalent to discarding the associated columns of $\Fbf^{(n)}$.
\end{IEEEproof}

The codeword $\phi(\wbf) \in \Ccal^{(n)}$ is therefore the sum of two linearly independent lattice points: $\phi_r(\wbf)$, which we call the {\em resolution information} and which encodes the first $k_r\log_2 p$ bits of the message, and $\phi_v(\wbf)$, which we call the {\em vestigial information} and which encodes the remaining $k_v \log_2 p$ bits. Furthermore, the decomposition is linear in the sense that the decomposition of sums of lattice points is the same as the sum of decompositions.
\begin{lemma}\label{lem:linear.decomposition}
	Let $\wbf_1$ and $\wbf_2$ be messages in $\mathbb{F}_p^k$, and let $\wbf = \wbf_1 \oplus \wbf_2$. Then
	\begin{equation}
		\phi_r(\wbf) = [\phi_r(\wbf_1) + \phi_r(\wbf_2)] \mod \Lambda_s^{(n)},
	\end{equation}
	and
	\begin{equation}
		\phi_v(\wbf) = [\phi_v(\wbf_1) + \phi_v(\wbf_2)] \mod \Lambda_s^{(n)}.
	\end{equation}
\end{lemma}
\begin{IEEEproof}
	This follows directly from the fact that $\phi$ is an isomorphism:
	\begin{align}
		\phi_r(\wbf) &= \phi_r(\wbf_1 \oplus \wbf_2) \\
		&= \phi(\wbf_1[1:k_r] \oplus \wbf_2[1:k_r]) \\
		&= [\phi(\wbf_1[1:k_r]) + \phi(\wbf_2[1:k_r])] \mod \Lambda_s^{(n)} \\
		&= [\phi_r(\wbf_1) + \phi_r(\wbf_2)] \mod \Lambda_s^{(n)}.
	\end{align}
	A similar argument holds for $\phi_v$.
\end{IEEEproof}

The preceding decomposition permits a lattice-coding instantiation of block Markov encoding. After the transmission of a lattice codeword, cooperating users can transmit the lower-rate resolution component. The intended receiver first decodes the resolution component and subtracts it from the received signal; the receiver then needs only to decode the lower-rate vestigial component. Although in this paper we apply the technique to compute-and-forward, it can be applied to a variety of relay systems. For example, in a related work \cite{song:IT11} an innovative lattice {\em list decoding} technique is proposed to show that lattice codes can achieve the capacity of the physically degraded three-terminal relay channel. Using our technique, it is straightforward to construct an alternative strategy that establishes the same result.

\section{Cooperative Computation: Encoding Strategy and Achievable Rate}\label{sect:achievable.rate}
In this section we describe our cooperative computation strategy and derive its achievable rate, which amounts to a proof of Theorem \ref{thm:subset}. Our approach is based on the lattice decomposition described in Section \ref{sect:lattice.subspaces}. Messages are communicated in two stages. In the first stage, each transmitter sends the lattice codeword corresponding to its message; this codeword is decoded by a subset of the other transmitters. In the second stage, transmitters cooperatively transmit the resolution component of the linear combinations desired at the receivers. Similarly, receivers decode in two stages. They first decode the resolution component, which they then subtract away from the received signal. Then they need only to decode the vestigial component, which is easier to decode due to its lower rate.

\begin{IEEEproof}[Proof of Theorem \ref{thm:subset}]
	Our proof goes in three parts: a description of the encoding scheme, a description of the decoding scheme, and an analysis of the probability of error.
	
	{\bf Encoding: } Each transmitter employs identical lattice codebooks\footnote{Technically we have a sequence of lattice codebooks indexed by $n$. In the interest of notational simplicity we drop the superscripts.} $\Ccal$ having rate $R_c$. The codebook $\Ccal$ decomposes into resolution and vestigial codebooks $\Ccal_r$ and $\Ccal_v$ which have respective rates $R_r$ and $R_v$. As noted in Section \ref{sect:lattice.subspaces}, we have $R_c = R_r + R_v$.
		
	Transmitters encode their $T$ messages over $T+1$ blocks as depicted in Table \ref{tab:bme}. At block $t$, each transmitter $l$ has a message $\wbf_l(t)$, which it encodes by mapping it to the corresponding codeword in $\Ccal$:
	\begin{equation}
		\lambda_l(t) = \phi(\wbf_l(t)).
	\end{equation}
	By Lemma \ref{lem:subspace.decomposition}, each lattice codeword can be decomposed by projecting onto the resolution and vestigial codebooks:
	\begin{align*}
		\lambda_{r,l}(t) &= \phi_r(\wbf_l(t)) \\
		\lambda_{v,l}(t) &= \phi_v(\wbf_l(t)).
	\end{align*}
	When transmitting the lattice codeword, each user {\em dithers} the lattice point over the shaping region; this ensures that the codebook looks approximately Gaussian as well as makes the codewords of each transmitter statistically independent\footnote{For further discussion of the need for dithers, see \cite{forney:Allerton04}}. We therefore define the effective codeword
	\begin{equation}
		\cbf_l(t) = [\lambda_l(t) + \tbf_l(t)] \mod \Lambda_s,
	\end{equation}
	where $\tbf_l(t)$ is a dither drawn randomly and uniformly over $\Vcal_s$, independent for each $1 \leq l \leq L$ and $1 \leq t \leq T$. Each receiver $m$ intends to recover the finite-field linear combination $\fbf_m(t) = \bigoplus_{l=1}^La_{lm} \wbf_m(t)$, which corresponds to the lattice point
	\begin{equation}
		\lambda_m(t) = \phi(\fbf(t)) = \left[\sum_{l=1}^L a_{lm} \lambda_l(t) \right] \mod \Lambda_s.
	\end{equation}
	
	As with the individual codewords, we can decompose $\lambda_m(t)$ into resolution and vestigial components:
	\begin{align}
		\lambda_{r,m}(t) &= \phi_r(\fbf_m(t)) \\
		\lambda_{v,m}(t) &= \phi_v(\fbf_m(t)).
	\end{align}
	The transmitters in $\mathcal{B}$ will cooperatively transmit $\lambda_{r,m}(t)$ to each receiver, again dithering the lattice point over $\Vcal_s$. The effective codeword is
	\begin{equation}
		\cbf_{r,m} = [\lambda_{r,m}(t) + \sbf_{m}(t)] \mod \Lambda_s,
	\end{equation}
	where, similar to before, $\sbf_{m}(t)$ is a dither drawn uniformly over $\Vcal_s$ and independent for each $1 \leq m \leq M$, and $1 \leq t \leq T$.

	At block $t=1$, each transmitter simply sends its own lattice codeword:
	\begin{equation}
		\xbf_l(1) = \sqrt{P}v_{l0}\cbf_{l}(t).
	\end{equation}	
	For subsequent blocks $2 \leq t \leq T$, each transmitter in $\mathcal{B}$ sends a combination of ``fresh'' information corresponding to its own message $\wbf_l(t)$ and resolution information corresponding to the messages sent in the previous time slot. Suppose that each transmitter in $\mathcal{B}$ has successfully decoded $\lambda_{l^\prime}(t-1)$ for each $l^\prime \neq l$. Then each transmitter in $\mathcal{B}$ can construct every $\lambda_m(t)$ and, by extension, every $\lambda_{r,m}(t)$. Every transmitter sends its own lattice codeword, and transmitters in $\mathcal{B}$ send the resolution components for each receiver:
	\begin{equation}
		\xbf_l(t) =
		\begin{cases}
			\sqrt{P}\left( v_{l0}\cbf_{l}(t) + \sum_{m=1}^M v_{lm} \cbf_{r,m}(t-1) \right), & \text{ for $l \in \mathcal{B}$} \\
			\sqrt{P}v_{l0}\cbf_{l}(t), &\text{ otherwise}
		\end{cases}.
	\end{equation}	
	Finally, at block $t = T+1$ there is no new fresh information for the transmitters to send. Each transmitter in $\mathcal{B}$ sends only the resolution information corresponding to block $T$, and the other transmitters send nothing:
	\begin{equation}
		\xbf_l(T+1) =
		\begin{cases}
			\sqrt{P}\sum_{m=1}^M v_{lm} \cbf_{r,b(l)m}(T), & \text{ for $l \in \mathcal{B}$} \\
			0 & \text{ otherwise}
		\end{cases}.
	\end{equation}
	
		\begin{table}[htb]
		\caption{Superposition Block Markov encoding for Theorem \ref{thm:subset}}\label{fig:bme.subset}
		\begin{center}
		\begin{tabular}{c|c|c|c|c}
			& $t=1$ & $t=2$ & $\cdots$ & $t=T+1$\\
			\hline
			$\xbf_1(t), 1 \in \mathcal{B}$ & $v_{10}\cbf_1(1)$ & $v_{10}\cbf_1(2) + \sum_{m=1}^M v_{1m}\cbf_{r,m}(1)$  & $\cdots$ & $\sum_{m=1}^M v_{1m}\cbf_{r,m}(T)$\\
			$\xbf_2(t), 2 \notin \mathcal{B}$& $v_{20}\cbf_2(1)$ & $v_{20}\cbf_2(2)$ & $\cdots$ & $0$\\
			$\vdots$ & $\vdots$ & $\vdots$ & $\vdots$ & $\vdots$ \\
			$\xbf_L(t), L \in \mathcal{B}$ & $v_{L0}\cbf_L(1)$ & $v_{L0}\cbf_L(2) + \sum_{m=1}^M v_{Lm}\cbf_{r,m}(1)$ & $\cdots$ &$ \sum_{m=1}^M v_{Lm}\cbf_{r,m}(T)$
		\end{tabular}
		\end{center}
		\label{tab:bme}
	\end{table}
	
	Note that, since $\Lambda_s$ has normalized second moment equal to unity, and since the dithers are independently and uniformly drawn from $\Vcal_s$, we have with high probability
	\begin{equation}
		\frac{1}{n}\norm{\xbf_l(t)}^2 \to P\sum_{m=0}^M v_{lm}^2 \leq P.
	\end{equation}
	Thus the transmit signals obey the average power constraint.
	
	{\bf Decoding:} Decoding proceeds in three stages. Each transmitter decodes the messages of every other transmitter, the receivers decode the resolution information send cooperatively by the clusters, and finally the receivers decode the vestigial information. Having decoded both components of the desired lattice point, the receiver can recover the desired linear function.
	
	At block $t=1$ each transmitter receives the superposition of all the other transmitters' signals, scaled by channel gains and corrupted by noise:
	\begin{align}
		\zbf_l(1) = \sqrt{P}\sum_{l^\prime \neq l}v_{l^\prime 0}g_{l^\prime l}\cbf_{l^\prime}(1) + \nbf_l(t).
	\end{align}
	Each transmitter forms estimates $\hat{\wbf}_{l^\prime l}(1)$ for every $l^\prime \neq l$ via typical sequence decoding: if there is a unique collection of messages jointly typical with the received signal, that collection is taken as the estimate; otherwise an error is declared. Note that in this case the transmitters do {\em not} employ lattice decoding.

	For blocks $2 \leq t \leq T$ the situation is similar. Each transmitter receives the superposition of other transmitters' signals, but in this case the received signals also contain resolution information:			\begin{equation}
		\zbf_l(t) = \sqrt{P}\left(\sum_{l^\prime \neq l}g_{l^\prime l}v_{l^\prime 0}\cbf_{l^\prime}(t) +  \sum_{l^\prime \in \mathcal{B}}\sum_{m=1}^Mg_{l^\prime l}v_{l^\prime m}\cbf_{r,m}(t-1)\right)+ \nbf_l(t).
	\end{equation}
	Supposing that each transmitter has successfully decoded the messages from block $t-1$, it knows the resolution information. It therefore can subtract this component out, resulting in the effective signal
	\begin{align}
		\zbf_l^\prime(t) &= \zbf_l(t) - \sqrt{P} \sum_{l^\prime \in \mathcal{B}}\sum_{m=1}^Mg_{l^\prime l}v_{l^\prime m}\cbf_{r,m}(t-1) \\
		&= \sqrt{P}\sum_{l^\prime \neq l}g_{l^\prime l}v_{l^\prime 0}\cbf_{l^\prime}(t) + \nbf_l(t)
	\end{align}
	Now, just as for $t=1$, each transmitter can form estimates $\hat{\wbf}_{l^\prime l}(t)$ of the other transmitters' messages via typical sequence decoding.
	
	Next we turn to the receivers. To decode the function $\fbf_m(t)$, each receiver first decodes the resolution information from the signal received in block $t+1$:
	\begin{multline}
		\ybf_m(t+1) = \sqrt{P}\sum_{l=1}^L h_{lm} v_{l0} \cbf_l(t+1) + \sqrt{P}\sum_{m^\prime \neq m}^M\sum_{l \in \mathcal{B}} h_{lm}v_{lm^\prime} \cbf_{l,m^\prime}(t) + \\ 
		\sqrt{P}\sum_{l \in \mathcal{B}} h_{lm}v_{lm}\cbf_{r,m}(t) + \nbf_m(t+1).
	\end{multline}
	Each receiver decodes the resolution information treating the interference---in this case the fresh information from each transmitter and the resolution information intended for other receivers---as noise. Each estimate $\hat{\lambda}_{r,m}(t)$ is formed via lattice decoding as outlined in \cite{erez:IT04}. The receivers first apply MMSE scaling to the incoming signal and subtract off the dither. Let
	\begin{equation}\label{eqn:resolution.noise}
		\nbf^\prime_m(t+1) =  \sqrt{P}\sum_{l=1}^L h_{lm} v_{l0} \cbf_l(t+1) + \sqrt{P}\sum_{m^\prime \neq m}^M\sum_{l \in \mathcal{B}} h_{lm}v_{lm^\prime} \cbf_{l,m^\prime}(t) + \nbf_m(t+1)
	\end{equation}
	be the sum of the interference and noise at receiver $m$. Then the scaled signal is
	\begin{align}
		\ybf^\prime_m(t+1) &= \left[ \gamma_m(t+1)\ybf_m(t+1) - \sbf_m(t+1) \right] \mod \Lambda_s \\
		&= \left[ \gamma_m(t+1)\sqrt{P}\sum_{l \in \mathcal{B}} h_{lm}v_{lm}\cbf_{r,m}(t) + \gamma \nbf^\prime_m(t+1) - \sbf_m(t+1) \right] \mod \Lambda_s \\
		&= \left[ \lambda_{r,m}(t+1) + \left(\gamma_m(t+1)\sqrt{P}\sum_{l \in \mathcal{B}} h_{lm}v_{lm} - 1\right)\cbf_{r,m}(t) + \gamma_m(t+1) \nbf^\prime_m(t+1) \right] \mod \Lambda_s \\
		&= \left[ \lambda_{r,m}(t+1) + \nbf_m^{\prime\prime}(t+1) \right] \mod \Lambda_s,
	\end{align}
	where
	\begin{equation}
		\nbf_m^{\prime\prime}(t+1) = \left(\gamma_m(t+1)\sqrt{P}\sum_{l \in \mathcal{B}} h_{lm}v_{lm} - 1\right)\cbf_{r,m}(t) + \gamma_m(t+1) \nbf^\prime_m(t+1)
	\end{equation}
	is the effective noise, including thermal noise, interference, and self-noise associated with MMSE scaling. Then, the estimate is formed by lattice quantization:
	\begin{equation}
		\hat{\lambda}_{r,m}(t) = Q_{\Lambda_r}(\ybf^\prime_m(t+1)).
	\end{equation}
	
	After decoding the resolution information, each receiver turns to $\ybf_m(t)$ to decode the vestigial component $\lambda_{v,m}(t)$. First, we note that, supposing that each receiver has successfully decoded the resolution information from the previous block, it can subtract that portion of the interference, yielding:
	\begin{align}
		\ybf_m^\prime(t) &= \ybf_m(t) - \sqrt{P}\sum_{l \in \mathcal{B}} h_{lm}v_{lm}\cbf_{r,m}(t-1) \\
		&= \sqrt{P}\sum_{l=1}^L h_{lm} v_{l0} \cbf_{l}(t) + \sqrt{P}\sum_{m^\prime \neq m}\sum_{l \in \mathcal{B}} h_{lm}v_{lm^\prime} \cbf_{r,m^\prime}(t-1) + \nbf_m(t).
	\end{align}
	Furthermore, supposing that the resolution information was decoded successfully, each receiver can subtract $\lambda_{r,m}(t)$ from the received signal modulo the shaping lattice. Finally, in preparation for lattice decoding, we apply MMSE scaling to the signal and subtract the dithers as in \cite{erez:IT04,nazer:IT11}. Let
	\begin{equation}\label{eqn:vestigial.noise}
		\nbf_m^\prime(t) = \sqrt{P}\sum_{m^\prime \neq m}\sum_{l \in \mathcal{B}} h_{lm}v_{lm^\prime} \cbf_{r,m^\prime}(t-1) + \nbf_m(t)
	\end{equation}
	be the sum of the interference and noise in $\ybf_m(t)$. The resulting signal is then
	\begin{align}
		\ybf^{\prime\prime}_m(t) &= \left[\alpha_m(t) \ybf^{\prime}_m(t) - \lambda_{v,m}(t) - \sum_{l=1}^L a_{lm}\tbf_l(t) \right] \mod \Lambda_s \\
		&= \left[ \sum_{l=1}^L (\alpha_m(t) \sqrt{P}h_{lm}v_{l0 }\cbf_l(t) - a_{lm}\tbf_l(t)) - \lambda_{r,m}(t) + \alpha_m(t)\nbf^\prime_m(t) \right] \mod \Lambda_s \\
		&= \left[ \sum_{l=1}^L a_{lm}(\cbf_l(t) - \tbf_l(t)) - \lambda_{r,m}(t) + \sum_{l=1}^L (\alpha_m(t)\sqrt{P}h_{lm}v_{lm} - a_{lm})\cbf_l(t) + \alpha_m(t)\nbf^\prime_m(t) \right] \mod \Lambda_s \\
		&= \left[ \lambda_{m}(t) - \lambda_{r,m}(t) + \sum_{l=1}^L (\alpha_m(t)\sqrt{P}h_{lm}v_{lm} - a_{lm})\cbf_l(t) + \alpha_m(t)\nbf^\prime_m(t) \right] \mod \Lambda_s \\
		&= \left[ \lambda_{v,m}(t) + \sum_{l=1}^L (\alpha_m(t)\sqrt{P}h_{lm}v_{lm} - a_{lm})\cbf_l(t) + \alpha_m(t)\nbf^\prime_m(t) \right] \mod \Lambda_s \\
		&= \left[ \lambda_{v,m}(t) + \nbf^{\prime\prime}_m(t) \right] \mod \Lambda_s, \label{eqn:vestigial.signal}
	\end{align}
	where
	\begin{equation}
		\nbf^{\prime\prime}(t) = \sum_{l=1}^L (\alpha_m(t)\sqrt{P}h_{lm}v_{lm} - a_{lm})\cbf_l(t) + \alpha_m(t)\nbf^\prime_m(t)
	\end{equation}
	is the effective noise, including thermal noise, interference from other transmitters and clusters, and self-noise associated with MMSE scaling. Each receiver decodes the estimate $\hat{\lambda}_{v,m}(t)$ by quantizing to the nearest point in $\Lambda_v$:
	\begin{equation}
		\hat{\lambda}_{v,m}(t) = Q_{\Lambda_v}(\ybf^{\prime\prime}_{m}(t)).
	\end{equation}
	
	Finally, having recovered both the resolution and vestigial components, each receiver constructs its estimate of the desired lattice codeword, from which it can recover the desired finite-field message:
	\begin{equation}
		\hat{\fbf}_m(t) = \phi^{-1}(\hat{\lambda}_{m}(t)) = \phi^{-1}\left(\left[\hat{\lambda}_{r,m}(t) + \hat{\lambda}_{v,m}(t) \right] \mod \Lambda_s\right).
	\end{equation}
	
		{\bf Probability of error:} An error occurs when (a) any of the transmitters in $\mathcal{B}$ fails to decode the other transmitters' messages, (b) any of the receivers fails to decode correctly the incoming resolution information, or (c) when any of the receivers fails to decode correctly the vestigial information associated with the desired lattice point. By the union bound, the probability of error follows			
	\begin{align}
		P_e &\leq \sum_{t=1}^T\sum_{m=1}^M\mathrm{Pr}\{\hat{\fbf}_m(t) \neq \fbf_m(t)\} \\
		&\leq \sum_{t=1}^T\sum_{l \in \mathcal{B}}\sum_{l^\prime \neq l}\mathrm{Pr}\{\hat{\wbf}_{l^\prime l}(t) \neq \wbf_{l^\prime}(t) \} + \sum_{t=1}^T\sum_{m=1}^M\mathrm{Pr}\{\hat{\lambda}_{r,m}(t) \neq \lambda_{r,m}(t)\} + \sum_{t=1}^T\sum_{m=1}^M \mathrm{Pr}\{\hat{\lambda}_{v,m}(t) \neq \lambda_{v,m}(t) \}. \label{eqn:error.probability}
	\end{align}
	
	Here we show that as long as the rates satisfy (\ref{eqn:subset}), each error term in (\ref{eqn:error.probability}) goes to zero exponentially. We start with the first summation. Each transmitter decodes the messages within its cluster via typical sequence decoding while treating all out-of-cluster interference as noise. By Lemma \ref{lem:lattice.to.gaussian} the joint mutual information between the transmit codewords $\cbf_l(t)$ and the receive signal $\zbf^\prime(t)$ approaches that of a Gaussian multiple-access channel with channel coefficients $g_{l^\prime l}v_{l^\prime 0}$, transmit power $P$, and unit noise power. Therefore, so long as
	\begin{equation}
		R < \min_{l \in \mathcal{B} }C_{\mathrm{mac}}(\mathbf{g}_l[1:l-1,l+1:L]\circ\mathbf{v}_0[1:l-1,l+1:L], P, 1),
	\end{equation}
	then $\mathrm{Pr}\{\hat{\wbf}_{l^\prime l}(t) \neq \wbf_{l^\prime}(t) \} \to 0$ exponentially for each $l$ and $l^\prime \neq l$.
	
	Next we turn to the resolution information. Here each receiver decodes $\lambda_{r,m}(t)$ via lattice decoding on $\ybf_m(t+1)$. In \cite{erez:IT04} it is shown that lattice decoding is sufficient to achieve the capacity of the Gaussian channel. From (\ref{eqn:resolution.noise}) we compute that the the interference power in $\nbf^\prime_m(t+1)$ is
	\begin{equation}
		I_{m,r} = \frac{1}{n}E[\norm{\nbf^\prime_m(t+1)}^2] = P\left(\norm{ \hbf_m \circ \vbf_0 }^2 + \sum_{m^\prime \neq m,0} |\hbf^T_m \vbf_m^\prime |^2\right).
	\end{equation}
	Similarly, we compute that the power of the resolution information in $\ybf_m(t+1)$ is $P|\hbf^T\vbf_m|^2$. Putting these together, we find that if
	\begin{equation}
		R_r < \min_{1 \leq m \leq M} \frac{1}{2} \log_2\left(1+ \frac{P|\hbf_m^T \vbf_m|^2}{1 + P(\norm{ \hbf_m \circ \vbf_0 }^2 + \sum_{m^\prime \neq m} |\hbf^T_m \vbf_m^\prime |^2)}\right),
	\end{equation}
	then $\mathrm{Pr}\{\hat{\lambda}_{r,m}(t) \neq \lambda_{r,m}(t)\} \to 0$ exponentially for each $m$.
	
	Finally, we look at the vestigial information. Here each receiver decodes $\lambda_{v,m}(t)$ by lattice decoding the sum of multiple incoming lattice points, so we borrow the main result from \cite{nazer:IT11}. We compute the interference power in (\ref{eqn:vestigial.noise}) to be
	\begin{equation}
		I_{m,v} = P\sum_{m^\prime \neq m} |\hbf_m^T\vbf_{m^\prime}|^2,
	\end{equation}
	and the effective channel gains in (\ref{eqn:vestigial.signal}) are $\hbf_m \circ \vbf_0$. Applying these to the rate in (\ref{eqn:non.cooperative.rate}), we find that if
	\begin{equation}
		R_v < \left [\frac{1}{2} \log_2(\norm{ P \hbf_m \circ \vbf_0}^2 + I_{m,v}) -
		\frac{1}{2}\log_2\left(\norm{\mathbf{a}_m}^2(1+I_{m,v}) + P\left(\norm{\mathbf{a}_m}^2\norm{\mathbf{h}_m \circ \mathbf{v}_0}^2 - |\mathbf{a}_m^T(\mathbf{h}_m \circ \mathbf{v}_0)|^2\right)\right)\right]^+,
	\end{equation}
	then $\mathrm{Pr}\{\hat{\lambda}_{v,m}(t) \neq \lambda_{v,m}(t) \} \to 0$ exponentially.
	
	Recall that $R_c = R_r + R_v$ and $R = \frac{TR_c}{T+1}$. Choosing $T$ arbitrarily large, we obtain the desired result.
\end{IEEEproof}

\section{Establishing the Diversity-multiplexing Tradeoff}\label{sect:DMT}
In this section we detail the signaling strategies that establish our diversity-multiplexing claims. We begin with the non-cooperative case. Our approach is to choose the steering vector $\mathbf{v}_0$ such that the effective channel vector is constant and has unit gain. It turns out that this approach fails approximately as often as does a single SISO link.
\begin{IEEEproof}[Proof of Theorem \ref{thm:non.cooperative.dmt}]
	First we prove the upper bound. For the non-cooperative case, it is shown in \cite[Theorem~13]{nazer:IT11} that the computation capacity is upper-bounded by
	\begin{align}
		C(\mathbf{H},P) &\leq \max_{\mathbf{A} \in \mathcal{A}} \min_{\substack{l,m \\ a_{lm} \neq 0}} \frac{1}{2} \log_2(1+Ph_{lm}^2) \\
		&\leq \frac{1}{2} \log_2(1 + Ph_{lm}^2),
	\end{align}
	where $l$ and $m$ can be chosen arbitrarily such that $a_{lm} \neq 0$. Then the computation capacity is bounded by the Shannon capacity of a single SISO link, which is proven in \cite{zheng:IT03} to have diversity-multiplexing tradeoff $d^*(r) = 1-r$. The compute-and-forward network therefore has DMT bounded by
	\begin{equation}
		d^*(r) \leq d^+_{\mathrm{nc}}(r) = 1-r.
	\end{equation}
	
	To prove the lower bound for $M=1$, we invoke the non-cooperative rate of Corollary \ref{cor:non.cooperative}, choosing $\vbf_0$ to align with the channels. For multiplexing gain $r$, we choose $\mathbf{a} = \mathbf{1}$ and $v_l^2 = P^{r-1}/h_l^2$, resulting in the achievable rate
	\begin{align}
		R(\mathbf{H},P) &= \frac{1}{2}\log_2(1+LP^r) - \frac{1}{2}\log_2(L) \\
		&= \frac{1}{2}\log_2\left(\frac{1+LP^r}{L}\right) \\
		&\geq \frac{1}{2}\log_2(P^r).
	\end{align}
	Outage occurs only when we cannot set $v_l$ to the specified value. Since we are constrained to have $v_l^2 \leq 1$, this occurs when $h_l^2 \leq P^{r-1}$. The probability of outage is therefore
	\begin{align}
		P_o \leq \mathrm{Pr}\left\{ \bigcup_{l=1}^L h_l \leq P^{r-1} \right\} \leq \sum_{l=1}^L \mathrm{Pr} \left\{ h_l^2 \leq P^{r-1} \right\} \approx LP^{r-1}.
	\end{align}
	Therefore, our scheme gives a diversity order at multiplexing gain $r$ of
	\begin{align}
		d^-_{\mathrm{nc}}(r) &= \lim_{P \to \infty} -\frac{\log(P_o)}{\log(P)} \\
		&\geq \lim_{P \to \infty} \frac{(1-r)\log(P) - \log(L)}{\log(P)} \\
		&= 1-r.
	\end{align}
	Since this matches the upper bound, the DMT is established.	
\end{IEEEproof}

Next we prove the results for the cooperative network. First we prove the upper bound.
\begin{IEEEproof}[Proof of Theorem \ref{thm:cooperative.dmt}]
	We invoke the MISO outer bound on the computation capacity from Theorem \ref{thm:miso.upper.bound}. The symmetric-rate capacity of the MISO broadcast channel is trivially upper bounded by the capacity of the single-user MISO link between the source and any destination. Thus the DMT is upper-bounded by that of a single $L$-antenna MISO link, which is shown in \cite{zheng:IT03} to be $d^*(r) = L(1-r)$. Thus the DMT of the cooperative compute-and-forward network is bounded by
	\begin{equation}
		d^*(r) \leq d^+(r) = L(1-r).
	\end{equation}
\end{IEEEproof}

Next we prove Theorem \ref{thm:cooperative.dmt}, for which we need first to establish an achievable rate using random codes and time-sharing.
\begin{lemma}\label{lem:random.coding}
	Let $\mathcal{B} \subset \{1, \cdots, L\}$. In the cooperative compute-and-forward network with $M=1$ receiver, the following computation rate is achievable:
	\begin{equation}\label{eqn:random.coding}
		R_{\mathrm{random}}(\mathbf{H},\mathbf{G},P) = \min\left\{\min_{l \in \mathcal{B}} \frac{1}{2}C_{\mathrm{mac}}(\mathbf{g}_l[1:l-1,l+1:L],P,1), \frac{1}{4}\log_2(1+P(\hbf_{\mathcal{B}}^T\mathbf{1})^2) \right\}.
	\end{equation}
	\begin{IEEEproof}
		The encoding scheme is simple, so we only sketch the proof. Divide the transmission into two equal time blocks. At the first block, each transmitter encodes and broadcasts its message using a random Gaussian codebook of power $P$. The transmitters in $\mathcal{B}$ decode the incoming messages using typical sequence decoding. This is nothing more than a Gaussian multiple-access channel, so decoding is successful as long as the rate is below the first term in (\ref{eqn:random.coding}). The multiple-access rate is cut in half due to time sharing.
				
		At the second block, the transmitters in $\mathcal{B}$ directly encode and broadcast the linear combination desired at the receiver, again using a random Gaussian codebook of power $P$.. The receiver decodes the desired function from the signal received in the second block only. This is equivalent to a MISO channel with equal beamformer weights, so decoding is successful as long as the rate is below the second term in (\ref{eqn:random.coding}). Again the MISO rate is cut in half due to time sharing.
	\end{IEEEproof}
\end{lemma}

Now we are in a position to prove Theorem \ref{thm:random.dmt}.
\begin{IEEEproof}[Proof of Theorem \ref{thm:random.dmt}]
	We construct an achievable scheme based on the strategy from Lemma \ref{lem:random.coding}. We allow $\mathcal{B}$ to vary according to the channel realizations, giving us the achievable rate
	\begin{align}
		R(\mathbf{H},\mathbf{G},P) &= \max_{\mathcal{B}} \min\left\{\min_{l \in \mathcal{B}} \frac{1}{2}C_{\mathrm{mac}}(\mathbf{g}_l[1:l-1,l+1:L],P,1), \frac{1}{4}\log_2(1+P(\hbf_{\mathcal{B}}^T\mathbf{1})^2) \right\} \\
		&\geq \max_{\mathcal{B}} \min\left\{\min_{l \in \mathcal{B}} \frac{1}{2}C_{\mathrm{mac}}(\mathbf{g}_l[1:l-1,l+1:L],P,1), \frac{1}{4}\log_2(1+P(\norm{\hbf_{\mathcal{B}}}^2) \right\}. \label{eqn:random.bound}
	\end{align}
	Let each rate term in (\ref{eqn:random.bound}) be denoted by $R_{\mathcal{B}}(\mathbf{H},\mathbf{G},P)$. Then define the event in which a particular cooperation modality fails:
	\begin{equation}
		\mathcal{O}_{\mathcal{B}} =  \left\{ R_{\mathcal{B}}(\Hbf,\Gbf,P) < \frac{r}{2}\log(P)\right\}.
	\end{equation}
	Outage occurs when each cooperation modality fails simultaneously:
	\begin{align}
		\mathcal{O} &= \bigcap_{\mathcal{B}}\mathcal{O}_{\mathcal{B}} \\
		&\subset \bigcap_{l=1}^L \mathcal{O}_{\{l\}}.
	\end{align}
	That is, we consider only the events in which a single transmitter decodes the messages. Each term in (\ref{eqn:random.bound}) has two components, the failure of either of which results in the failure of the cooperation modality. Therefore, define two events: $\mathcal{C}_{l}$, the event that transmitter $l$ fails to decode the other transmitters' messages, and $\mathcal{N}_l$, the event that, even if transmitter $l$ decodes successfully, the receiver fails to decode the linear function. The first event can be expressed as
	\begin{align}
		\mathcal{C}_l &=  \left\{ \frac{1}{2}C_{\mathrm{mac}}(\mathbf{g}_{l},P,1) < \frac{r}{2}\log(P) \right\} \\
		&=  \bigcup_{\mathcal{L} \subset \{1,\cdots,L\} \setminus \{ l \}} \left\{ \frac{1}{4|\mathcal{L}|}\log_2\left(1+ P\sum_{l^\prime \in \mathcal{L}}g_{l^\prime l}^2\right) < \frac{r}{2}\log(P) \right\} \\
		&\approx  \bigcup_{\mathcal{L} \subset \{1,\cdots,L\} \setminus \{ l \}} \left\{ \sum_{l^\prime \in \mathcal{L}}g_{l^\prime l}^2 < P^{2|\mathcal{L}|r - 1} \right\} \\
		&\subset \bigcup_{\mathcal{L} \subset \{1,\cdots,L\} \setminus \{ l \}} \left\{ \bigcap_{l^\prime \in \mathcal{L}} \left\{ g_{l^\prime l}^2 < P^{2|\mathcal{L}|r - 1} \right\}\right\}.
	\end{align}
	The second event can be expressed as
	\begin{align}
		\mathcal{N}_l &= \left\{ \frac{1}{4}\log_2\left(1+P h_l^2\right) < \frac{r}{2}\log_2(P) \right\} \\
		&\approx \left\{h_l^2 < P^{2r-1} \right\}.
	\end{align}
	Since each cooperation modality involves a different set of channel coefficients, the failure events $\mathcal{O}_l$ are independent. Therefore we can bound the outage probability by
	\begin{align}
		\mathrm{Pr}(\mathcal{O}) &\leq \prod_{l=1}^L \mathrm{Pr}(\mathcal{C}_l \cup \mathcal{N}_l) \\
		&\lesssim \prod_{l=1}^L \left( \sum_{\mathcal{L}}\prod_{l^\prime \in \mathcal{L}}\mathrm{Pr}(g_{l^\prime l}^2 < P^{2|\mathcal{L}|r-1}) + \mathrm{Pr}( h_l^2 < P^{2r-1} ) \right) \\
		&\approx \left(\sum_{|\mathcal{L}|=1}^{L-1} (P^{2|\mathcal{L}|r-1})^{|\mathcal{L}|} + P^{2r-1} \right)^L \\
		&\approx \left(\max_{1 \leq |\mathcal{L}| \leq L-1} P^{|\mathcal{L}|(2|\mathcal{L}|r-1)} + P^{2r-1}\right)^L. \label{eqn:random.quadratics}
	\end{align}
	To find the terms with the largest error exponent, we need to find the value of $|\mathcal{L}|$ that maximizes the quadratics in (\ref{eqn:random.quadratics}). For instance, it is clear that for $r=0$ the maximizer is $|\mathcal{L}|=1$. In general, since the quadratics in question are positive, the maximizer is either $|\mathcal{L}|=1$ or $|\mathcal{L}|=L-1$. This gives us
	\begin{align}
		\mathrm{Pr}(\mathcal{O}) &\lesssim \left(\max\left\{ P^{(L-1)(2(L-1)r-1)},P^{2r-1} \right\}  + P^{2r-1}\right)^L \\
		&\approx   \left(\max\left\{ P^{(L-1)(2(L-1)r-1)},P^{2r-1} \right\}\right)^L. \label{eqn:final.random.probabilities}
	\end{align}
		Finally, plugging (\ref{eqn:final.random.probabilities}) into the definition of the DMT, we get
	\begin{align}
		d^*(r) &= \lim_{P \to \infty} \frac{\log(\mathrm{Pr}(\mathcal{O}))}{\log(P)} \\
		&\geq L \min\{1-2r, (L-1)(1-2(L-1)r)\}.
	\end{align}
\end{IEEEproof}

Next we prove the lower bound achieved by cooperative lattice coding. Tuning $\mathbf{v}_0$ to the channel vector $\mathbf{h}$ makes it too difficult for transmitters to decode others' messages, so we take $\mathbf{v}_0 = P^{-x}$, where $x$ depends on the multiplexing gain. Increasing $\mathbf{v}_0$ decreases the error probability at the transmitters, but it increases the Cauchy-Schwarz penalty and therefore the error probability at the receiver. For $r=0$, choosing $x=1/2$ is optimal, which gives us a diversity gain of approximately $1/2$ for each transmitter.
\begin{IEEEproof}[Proof of Theorem \ref{thm:lattice.dmt}]
	The proof follows a similar outline to that of Theorem \ref{thm:random.dmt}, except that we use the rates proved in Theorem \ref{thm:subset} using lattice codes. Again we allow the subset of cooperating users $\mathcal{B}$ to vary according to the channel realizations, and we choose $\mathbf{a} = \mathbf{1}$, resulting in the following achievable rate
	\begin{multline}\label{eqn:DMT.rates}
		R(\mathbf{H},\mathbf{G},P) = \max_{\mathcal{B}} \min\bigg\{ \min_{l \in \mathcal{B}}C_{\mathrm{mac}}(\mathbf{g}_l[1:l-1,l+1:L] \circ \mathbf{v}_0[1:l-1,l+1:L], P, 1),\\
			\bigg[\frac{1}{2}\log_2(1+P(\norm{\mathbf{h}\circ \mathbf{v}_0}^2 + |\mathbf{h}^T\mathbf{v}_1|^2)) -
			\frac{1}{2}\log_2\left(L + P\left(L\norm{\mathbf{h} \circ \mathbf{v}_0}^2 - |\mathbf{1}^T(\mathbf{h} \circ \mathbf{v}_0)|^2\right)\right)\bigg]^+ \bigg\}.
	\end{multline}
	Similar to before, we let each term in (\ref{eqn:DMT.rates}) be denoted by $R_{\mathcal{B}}(\mathbf{H},\mathbf{G},P)$ and define the events corresponding to the failure of each cooperation modality:
	\begin{equation}
		\mathcal{O}_{\mathcal{B}} =  \left\{ R_{\mathcal{B}}(\Hbf,\Gbf,P) < \frac{r}{2}\log(P)\right\}.
	\end{equation}
	Outage occurs when each cooperation modality fails simultaneously:
	\begin{align}
		\mathcal{O} &= \bigcap_{\mathcal{B}} \mathcal{O}_{\mathcal{B}} \\
		&\subset \mathcal{O}_{\{1,\dots,L \}} \cap \bigcap_{l=1}^L \mathcal{O}_{\{l\}} \cap \mathcal{O}_{\emptyset}. \label{eqn:error.events}
	\end{align}
	Here we consider the events in which {\em all} transmitters cooperate, in which $L-2$ individual transmitters cooperate, and in which no one cooperates. When $\mathcal{B} = \emptyset$, we use the strategy outlined in the proof of Theorem \ref{thm:non.cooperative.dmt}, choosing $v_l^2 = P^{r-1}/h_l^2$. Following that line of analysis, the non-cooperative modality fails only when every channel gain is too low:
	\begin{equation}
		\mathcal{O}_\emptyset \subset \left\{ \bigcup_{l=1}^L h_l < P^{r-1} \right\}.
	\end{equation}
	
	For $\mathcal{B} \neq \emptyset$, we choose $v_{l0} = P^{-x_\mathcal{B}/2}$ for every $l$, and $v_{l 1} = 1-P^{-x_\mathcal{B}/2}$ for every $l \in \mathcal{B}$; otherwise $v_{1l}=0$. Using this, we can bound the rate as follows:	
	\begin{multline}
		R_{\mathcal{B}}(\mathbf{H},\mathbf{G},P) \geq \min\bigg\{ \min_{l \in \mathcal{B}}C_{\mathrm{mac}}(\mathbf{g}_{l}[1:l-1,l+1:L],P^{1-x_\mathcal{B}},1),\\
			\bigg[\frac{1}{2}\log_2\left(1+P\sum_{l \in \mathcal{B}} |h_l|^2\right) - \frac{1}{2}\log_2\left(L + P^{1-x_\mathcal{B}}\left(L\norm{\mathbf{h}}^2 -\norm{\mathbf{h}}^2\right)\right)\bigg]^+ \bigg\}.
	\end{multline}
	For large $P$, we get
	\begin{equation}\label{eqn:subset.rate}
		R_{\mathcal{B}}(\mathbf{H},\mathbf{G},P) \geq \min\left\{ \min_{l \in \mathcal{B}}C_{\mathrm{mac}}(\mathbf{g}_l[1:l-1,l+1:L],P^{1-x_\mathcal{B}},1),
			\frac{1}{2}\log_2\left(\frac{P^{x_{\mathcal{B}}}\sum_{l \in \mathcal{B}} |h_l|^2}{(L-1)\norm{\mathbf{h}}^2}\right) \right\}.
	\end{equation}
	As before we define events corresponding to the failure of either term in (\ref{eqn:subset.rate}): $\mathcal{C}_\mathcal{B}$, the event that the transmitters in $\mathcal{B}$ fail to decode the other transmitters' messages, and $\mathcal{N}_\mathcal{B}$, the event that, even if the transmitters decode each other properly, the receiver fails to decode its linear function at the required rate. The first event can be expressed as
	\begin{align}
		\mathcal{C}_\mathcal{B} &=  \bigcup_{l \in \mathcal{B}} \left\{ C_{\mathrm{mac}}(\mathbf{g}_l[1:l-1,l+1:L],P^{1-x_\mathcal{B}},1) < \frac{r}{2}\log(P) \right\} \\
		&=  \bigcup_{l \in \mathcal{B}} \bigcup_{\mathcal{L} \subset \{1,\cdots,L\} \setminus \{ l \}} \left\{ \frac{1}{2|\mathcal{L}|}\log_2\left(1+ P^{1-x_\mathcal{B}}\sum_{l^\prime \in \mathcal{L}}g_{l^\prime l}^2\right) < \frac{r}{2}\log(P) \right\} \\
		&\approx  \bigcup_{l \in \mathcal{B}} \bigcup_{\mathcal{L} \subset \{1,\cdots,L\} \setminus \{ l \}} \left\{ \sum_{l^\prime \in \mathcal{L}}g_{l^\prime l}^2 < P^{|\mathcal{L}|r + x_\mathcal{B} - 1} \right\} \\
		&\subset \bigcup_{l \in \mathcal{B}} \bigcup_{\mathcal{L} \subset \{1,\cdots,L\} \setminus \{ l \}} \left\{\bigcap_{l^\prime \in \mathcal{L}} \left\{ g_{l^\prime l}^2 < P^{|\mathcal{L}|r + x_\mathcal{B} - 1} \right\}\right\}.
		\end{align}
	For $\mathcal{B} = \{1,\cdots,L\}$, the second event can be expressed as
	\begin{align}
		\mathcal{N}_{\{1,\cdots,L\}} &= \left\{ \frac{1}{2}\log_2\left(\frac{P^{x_{\mathcal{B}}}\norm{\hbf}^2}{(L-1)\norm{\mathbf{h}}^2}\right) < \frac{r}{2}\log_2(P) \right\} \\
		&= \left\{ P^{x_{\{1,\cdots,L\}}} < (L-1)P^r \right\}. \label{eqn:empty.set}
	\end{align}
	Based on (\ref{eqn:empty.set}), we choose $x_{\{1,\cdots,L\}} = r + \epsilon$ for any $\epsilon>0$. As $P \to \infty$, this forces $\mathcal{N}_{\{1,\cdots,L\}} \to \emptyset$ deterministically. 
	For $\mathcal{B} = \{l\}$, we can express the second event as
	\begin{align}
		\mathcal{N}_{\{ l \}} &= \left\{ \frac{1}{2}\log_2\left(\frac{P^{x_{\{l\}}} h_l^2}{(L-1)\norm{\mathbf{h}}^2}\right) < \frac{r}{2}\log_2(P) \right\} \\
		&= \left\{ \frac{h_l^2}{(L-1)\norm{\mathbf{h}}^2} < P^{r-x_{\{l\}}} \right\} \\
		&\subset \left\{h_l^2 < P^{r-x_{\{l\}}-\epsilon} \right\} \cup \left\{\norm{\mathbf{h}}^2 \geq \frac{P^{\epsilon}}{L-1} \right\} \\
		&\subset \bigcap_{l \in \mathcal{B}} \left\{h_l^2 < P^{r-x_{\{l\}}-\epsilon} \right\} \cup \left\{\norm{\mathbf{h}}^2 \geq \frac{P^{\epsilon}}{L-1} \right\}.
	\end{align}
	Combining the above with (\ref{eqn:error.events}), we get
	\begin{multline}\label{eqn:all.the.events}
		\mathcal{O} \subset    \left[\left( \bigcup_{l \in \{1,\cdots,L \}} \bigcup_{\mathcal{L} \subset \{1,\cdots,L\} \setminus \{ l \}} \bigcap_{l^\prime \in \mathcal{L}} \left\{ g_{l^\prime l}^2 < P^{|\mathcal{L}|r + r+\epsilon - 1} \right\} \right ) \right]  \cap \\
		\bigcap_{l \in \{1,\cdots,L \}} \left[  \left(  \bigcup_{\mathcal{L} \subset \{1,\cdots,L\} \setminus \{ l \}} \bigcap_{l^\prime \in \mathcal{L}} \left\{ g_{l^\prime l}^2 < P^{|\mathcal{L}|r + x_{\{l\}} - 1} \right\} \right) \cup \left(\left\{h_l^2 < P^{r-x_{\{l\}}-\epsilon} \right\} \right)\right] \cap \\
		\left\{ \bigcup_{l=1}^L h_l^2 < P^{r-1} \right\} \cup  \left\{\norm{\mathbf{h}}^2 \geq \frac{P^{\epsilon}}{L-1} \right\}.
	\end{multline}
	Equation (\ref{eqn:all.the.events}) contains too many terms to enumerate in full. Since we are concerned with asymptotic behavior, we need only look at the term with the highest error exponent. This term contains one channel failure in $\mathcal{C}_{\{1,\cdots,L\}}$, $L-2$ failures in $\mathcal{C}_{\{l\}} \cap \mathcal{N}_{\{l\}}$, and one failure in $\mathcal{N}_\emptyset$. The final error event, in which $\norm{\hbf}^2$ is too large, has negligible contribution to the error probability. Combining these, we get
	\begin{multline}\label{eqn:some.probabilities}
		\mathrm{Pr}(\mathcal{O}) \lesssim \mathrm{Pr} \left( \bigcup_{\mathcal{L} \subset \{1,\cdots,L\} \setminus \{ 1 \}} \bigcap_{l^\prime \in \mathcal{L}} \left\{ g_{l^\prime 1}^2 < P^{|\mathcal{L}|r + r+\epsilon - 1} \right\} \right) \mathrm{Pr}\left( \left\{ h_2^2 < P^{r-1}\right\} \right) \times \\
		\prod_{l=3}^L\mathrm{Pr}\left( \bigcup_{\mathcal{L} \subset \{1,\cdots,L\} \setminus \{ l \}} \bigcap_{l^\prime \in \mathcal{L}} \left\{ g_{l^\prime l}^2 < P^{|\mathcal{L}|r + x_{\{l\}} - 1} \right\}
		\cup \left\{|h_l|^2 < P^{r-x_{\{l\}}-\epsilon} \right\}
		\right).
	\end{multline}
	Since each term in (\ref{eqn:some.probabilities}) is independent, we can evaluate the probabilities separately, yielding
	\begin{align}
		\mathrm{Pr}(\mathcal{O}) &\lesssim \left(\sum_{\Lcal \subset \{1,\cdots,L\} \setminus \{1\}} (P^{|\Lcal|r+x+\epsilon-1})^{|\Lcal|}\right)(P^{r-1})\left(\sum_{\Lcal \subset \{1,\cdots,L\} \setminus \{3\}} (P^{|\Lcal|r - x-1})^{|\Lcal|} + P^{r-x-\epsilon}\right)^{L-2} \\
		&= P^{r-1}\left(\sum_{\Lcal \subset \{1,\cdots,L\} \setminus \{1\}} (P^{|\Lcal|(|\Lcal|r+x+\epsilon-1)}\right)\left(\sum_{\Lcal \subset \{1,\cdots,L\} \setminus \{3\}} (P^{|\Lcal|(|\Lcal|r - x-1)}) + P^{r-x-\epsilon}\right)^{L-2}, \label{eqn:more.probabilities}
	\end{align}
	where we have chosen $x_{\{l\}}=x$ for every $l$. Similar to the proof Theorem \ref{thm:random.dmt}, the maximizer of the quadratics in (\ref{eqn:more.probabilities}) is either $|\Lcal|=1$ or $|\Lcal|=L-1$. This gives us
	\begin{align}\label{eqn:final.probabilities}
		\mathrm{Pr}(\mathcal{O}) &\lesssim P^{r-1}\left(\max\left\{P^{2r-1+\epsilon},P^{(1-L)(1-rL) + (L-1)\epsilon}\right\}\right)\left(\max_x\min\left\{P^{r+x-1},P^{(1-L)(1-(L-1)r-x)},P^{r-x-\epsilon} \right\} \right)^{L-2}.
	\end{align}
	Finally, plugging (\ref{eqn:final.probabilities}) into the definition of the DMT, taking the supremum over all $\epsilon > 0$, and taking the maximum over all $x$ yields
	\begin{align}
		d^-_{\mathrm{c}}(r) &= \lim_{P \to \infty} \frac{\log(\mathrm{Pr}(\mathcal{O}))}{\log(P)} \\
		&\geq 1-r + \min\{[1-2r]^+,[(L-1)(1-rL)]^+ \} + \notag \\
		&\quad\quad\quad\quad \max_{0 \leq x \leq 1}(L-2) \min\{[1-x-r]^+,[(L-1)(1-(L-1)r-x)]^+,[x-r]^+ \}.
	\end{align}
\end{IEEEproof}

Although their proofs are similar, the strategies Theorems \ref{thm:random.dmt} and \ref{thm:lattice.dmt} achieve very different diversity-multiplexing tradeoffs. With random coding, transmitters decode and cooperatively transmit at separate blocks; such time division enables full diversity, but it leads to poor multiplexing performance. With lattice coding, on the other hand, we need to balance transmit power in order to ameliorate the effects of signal misalignment; the balance costs us diversity gain, but the multiplexing performance is improved. The overall message is that transmit cooperation improves diversity {\em and} multiplexing for compute-and-forward, and as we saw in Figure \ref{fig:dmt.plot} the two approaches combined achieve the corner points of the DMT region.

\section{Numerical Examples}\label{sect:simluations}
In this section we examine a few example scenarios in which to demonstrate the benefits of our approach.
\begin{example}\label{ex:inter.channel}
	The first example, depicted in Figure \ref{fig:inter.channel}, comprises $L=2$ transmitters and a single receiver. The channels are symmetric, with the forward coefficients constant $h_1=h_2=1$ and the inter-transmitter coefficients a variable $g_{12}=g_{21}=g$, which we vary such that the gain $g^2$ ranges between $-10$dB and $30$dB. We set the transmit SNR at $P=10$dB. Since the channel gains are symmetric, either both transmitters can decode the other's message or neither of them can; therefore we choose either $\mathcal{B} = \{1,2\}$ or $\mathcal{B} = \emptyset$ for cooperative computation. Similarly, by symmetry it is easy to see that the optimal choice for the linear function is $\mathbf{a} = (1, 1)^T$ and that the optimal steering vectors $\mathbf{v}_0$ and $\mathbf{v}_1$ are constant. We find the optimal tradeoff between $\mathbf{v}_0$ and $\mathbf{v}_1$ numerically.

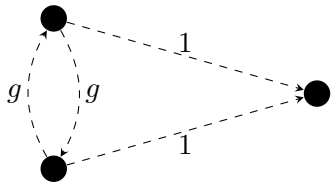
\begin{figure}[htb]
	\centering
\begin{tikzpicture}
		[scale=0.5,>=stealth,every node/.style={inner sep=0, minimum size=10}]
		\node[circle,fill] (x1) at (-2,2) {};
		\node[circle,fill] (x2) at (-2,-2) {}; 
				

		
		\draw [dashed,->] (x1) to [bend left]  node[right] {$g$} (x2);
		\draw [dashed,->] (x2) to [bend left] node[left] {$g$} (x1);


		\node[circle,fill] (y1) at (5,0) {};
		
		
		
		\draw [dashed,->] (x1) to node[above] {$1$} (y1);
		\draw [dashed,->] (x2) to node[below] {$1$} (y1);
		
		

\end{tikzpicture}
	\caption{Example \ref{ex:inter.channel}: A two-by-one computation network with symmetric channel gains.}
	\label{fig:inter.channel}
\end{figure}
	
	In Figure \ref{fig:inter.channel.plot} we plot the achievable rate of our cooperative scheme against the upper bound of Theorem \ref{thm:relay.upper.bound}, using the Nazer-Gastpar rate of (\ref{eqn:non.cooperative.rate}) as a baseline. The trends are easy to appreciate. When the channels between transmitters are weak, decoding each other's messages is too difficult, and the cooperative rate collapses to (\ref{eqn:non.cooperative.rate}). As the inter-transmitter gains become stronger it becomes easier for the transmitters to decode, and cooperation can improve the computation rate and eventually approaches the upper bound. We note a ``dimple'' in the cooperative rate as $g^2$ becomes large. For sufficiently large $g^2$, the optimal strategy is to turn the steering vector $\mathbf{v}_0$ down far enough that the Nazer-Gastpar component of the cooperative rate is zero, meaning that only the jointly-encoded resolution information carries information to the receiver. At this value of $g^2$ we see the dimple, after which the rate quickly converges on the upper bound.

	\begin{figure}[htb]
		\centering
\begin{tikzpicture}
	\pgfplotsset{every axis legend/.append style={at={(0.95,0.05)}, anchor=south east}} 

	\begin{axis}[
	xlabel={$g^2_{12} = g^2_{21}$ (dB)},
	ylabel={Computation rate (bits per channel use)},
	ymin = 1,
	grid=major,
	axis x line=bottom,
	axis y line=left,
	height=200pt,
	width=250pt
	]

		\addplot[smooth,color=red,very thick,solid] file {trans/noncooperative.inter.channel};
		\addlegendentry{\small Non-cooperative computation};
		
		\addplot[smooth,color=blue,very thick,dashed] file {trans/cooperative.inter.channel};
		\addlegendentry{\small Cooperative computation};
		
		\addplot[smooth,color=green!50!black, very thick,dotted] file {trans/upper.inter.channel};
		\addlegendentry{\small Upper bound (Theorem \ref{thm:relay.upper.bound})};

	\end{axis}
\end{tikzpicture}
		\caption{Achievable rates as a function of inter-transmitter channel gains for Example \ref{ex:inter.channel}.}
		\label{fig:inter.channel.plot}
	\end{figure}
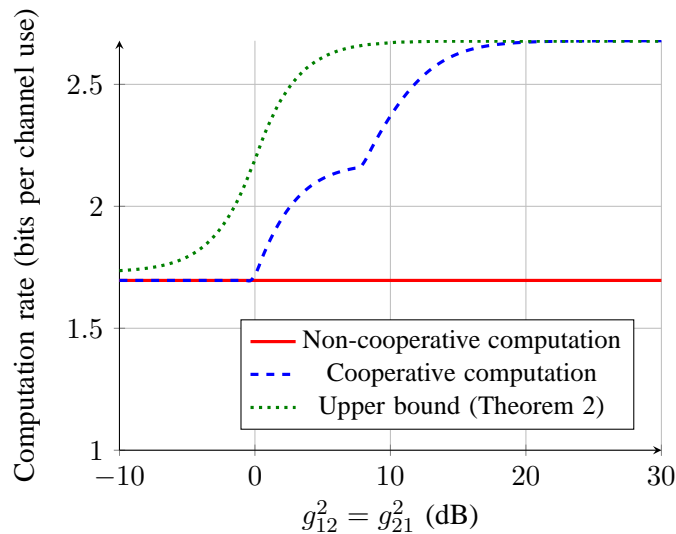	
\end{example}

\begin{example}\label{ex:path.loss}
Next we examine a scenario in which channel gains are chosen randomly, as depicted in Figure \ref{fig:simulation.model}. We place a single receiver at the origin and place $L=3$ transmitters randomly and uniformly on a segment of the circle having specified arclength. From the geometric configuration of the network, we compute channel magnitudes according to a path-loss model:
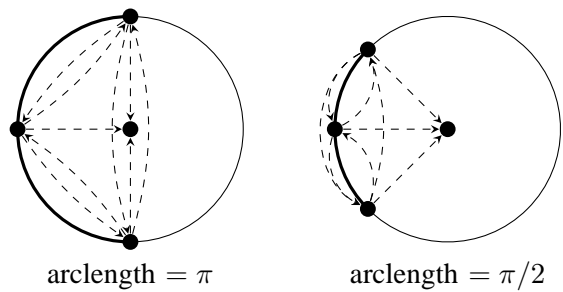
\begin{figure}[htb]
\centering
\begin{tikzpicture}
		[scale=1.5, every node/.style={inner sep=0, minimum size=6},>=stealth]

		\node[circle, fill] (d) at (0,0) {};
		
		\node[circle, fill] (u1) at (-90:-1) {};
		\node[circle, fill] (u2) at (0:-1) {};
		\node[circle, fill] (u3) at (90:-1) {};
		
		\draw[very thick] (-90:-1) arc (-90:90:-1);
		\draw (90:-1) arc (90:360:-1);
		
		\draw [dashed,->] (u1) to (d);
		\draw [dashed,->] (u2) to (d);
		\draw [dashed,->] (u3) to (d);
		
		\draw [dashed,->] (u1) to [bend right = 10] (u2);
		\draw [dashed,->] (u2) to [bend right = 10] (u1);
		
		\draw [dashed,->] (u1) to [bend right = 15] (u3);
		\draw [dashed,->] (u3) to [bend right = 15] (u1);
		
		\draw [dashed,->] (u2) to [bend right = 10] (u3);
		\draw [dashed,->] (u3) to [bend right = 10] (u2);	
		
		\node at (0,-1.3) {arclength $=\pi$};
		
		\begin{scope}[xshift=80]
			\node[circle, fill] (d) at (0,0) {};
		
			\node[circle, fill] (u1) at (-45:-1) {};
			\node[circle, fill] (u2) at (0:-1) {};
			\node[circle, fill] (u3) at (45:-1) {};
			
			\draw[very thick] (-45:-1) arc (-45:45:-1);
			\draw (45:-1) arc (45:360:-1);
			
			\draw [dashed,->] (u1) to (d);
			\draw [dashed,->] (u2) to (d);
			\draw [dashed,->] (u3) to (d);
			
			\draw [dashed,->] (u1) to [bend right = 40] (u2);
			\draw [dashed,->] (u2) to [bend right = 40] (u1);
			
			\draw [dashed,->] (u1) to [bend right = 63] (u3);
			\draw [dashed,->] (u3) to [bend right = 18] (u1);
			
			\draw [dashed,->] (u2) to [bend right = 40] (u3);
			\draw [dashed,->] (u3) to [bend right = 40] (u2);	
			
			\node at (0,-1.3) {arclength $=\pi/2$};
		\end{scope}

\end{tikzpicture}
	\caption{Example \ref{ex:path.loss}: Three users are placed along a segment of the unit circle, while the receiver is placed at the origin. \label{fig:simulation.model}}
\end{figure}
\begin{equation*}
	g_{ij} = \sqrt{\frac{1}{d(i,j)^\alpha}}, \quad h_i = \sqrt{\frac{1}{d(i,0)^\alpha}},
\end{equation*}
where $d(i,j)$ is the Euclidean distance between users $i$ and $j$ and. We choose $P=10$dB and a path-loss exponent of $\alpha=4$.

For each realization we calculate the cooperative computation rate. Since the gains from transmitters to receiver are equal, $\mathbf{a} = (1,1,1)^T$ is the optimal choice. The steering vectors and the clusters are optimized numerically. We run 500 simulations each for arclengths varying from 0 to $\pi$, and plot the average computation rates in Figure \ref{fig:rates}. Again the trends are easy to appreciate. Cooperation offers the greatest improvement when transmitters are close together. Even as we spread transmitters further apart, on average enough transmitters can cooperate that our approach garners a noticeable improvement.

\begin{figure}[htb]
	\centering
\begin{tikzpicture}[scale=0.88]
	
	\begin{axis}[
	xlabel={Distance between users},
	ylabel={Computation rate (bits per channel use)},
	axis x line=bottom,
	axis y line=left,
	ymin=0,
	legend style={at={(0.35,0.15)},
	anchor=south west},
	grid=major,
	xtick = \empty,
	extra x ticks = {0, 0.785,1.5708,2.356,3.14159},
	extra x tick labels = {$0$,$\pi/4$, $\pi/2$,$\pi/4$,$\pi$}
	]

		\addplot[smooth,color=red,thick,solid] file {trans/noncooperative.rates};
		\addlegendentry{\small Non-cooperative computation};
		
		\addplot[smooth,color=blue,thick,dashed] file {trans/selection.rates};
		\addlegendentry{\small Cooperative computation};
		

	\end{axis}
\end{tikzpicture}
	\caption{Average computation rate vs. angle between transmitters for Example \ref{ex:path.loss}.}
	\label{fig:rates}
\end{figure}
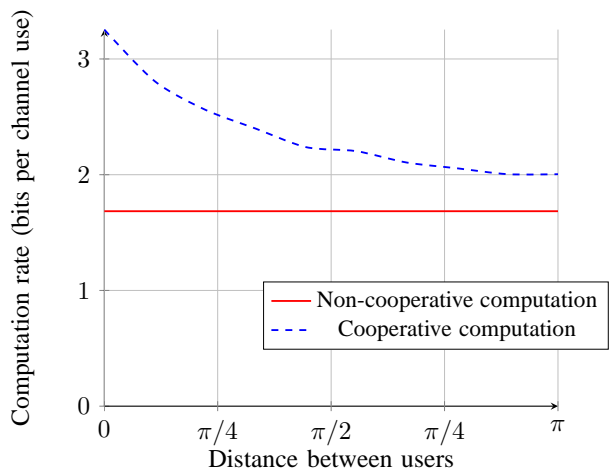

\end{example}

\begin{example}\label{ex:two.by.one}
In the next example we examine the variation in cooperative computation rate with respect to the channel gain between transmitters and receivers. As depicted in Figure \ref{fig:two.by.one}, we again have $L=2$ and $M=1$, but now we set every channel gain to unity except for $h_{12}$. Since the channels between transmitters and receiver are not symmetric, we cannot take $\mathbf{a} = (1,1)^T$ or $\mathbf{v}_0$ and $\mathbf{v}_1$ to be constant. Instead, we iterate manually through possible choices of $\mathbf{a}$ and numerically optimize over the set $\mathcal{B}$ of cooperating transmitters and the steering vectors $\mathbf{v}_0$ and $\mathbf{v}_1$.

\begin{figure}[htb]
	\centering
\begin{tikzpicture}
		[scale=0.5,>=stealth,every node/.style={inner sep=0, minimum size=10}]
		\node[circle,fill] (x1) at (-2,2) {};
		\node[circle,fill] (x2) at (-2,-2) {}; 
				

		
		\draw [dashed,->] (x1) to [bend left]  node[right] {$1$} (x2);
		\draw [dashed,->] (x2) to [bend left] node[left] {$1$} (x1);


		\node[circle,fill] (y1) at (5,0) {};
		
		
		
		\draw [dashed,->] (x1) to node[above] {$1$} (y1);
		\draw [dashed,->] (x2) to node[below] {$h_{21}$} (y1);
		
		

\end{tikzpicture}
	\caption{Example \ref{ex:two.by.one}: A two-by-one computation network with asymmetric channel gains.}
	\label{fig:two.by.one}
\end{figure}
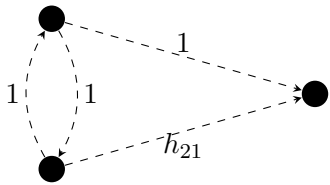

In Figure \ref{fig:two.by.one.rates} we plot the cooperative rate alongside (\ref{eqn:non.cooperative.rate}) for a variety of transmit signal-to-noise ratios $P$. We make a few observations. First, the non-cooperative rate is low for $h_{21}$ near to zero. Since we require the function to contain elements from both transmitters' messages, it becomes difficult for the receiver to decode such a function. In the cooperative case, however, the rates do not fall, since transmitter 1 can decode $\wbf_2$ and transmit the desired function to the receiver. This result hints at the diversity gains inherent to the cooperative approach; even when one link fails, successful computation is possible.

Furthermore, in the cooperative case we get the full multiplexing gain as the SNR becomes large. In the non-cooperative case we observe ``peaks''; these correspond to rational channel gains with low denominator. The further $h_{21}$ is from a low-denominator rational, the harder it is to align the function with the channels and the higher the Cauchy-Schwarz penalty in (\ref{eqn:non.cooperative.rate}). However, we can always choose $\vbf_0$ such that the equivalent channel vector is rational, allowing us to completely eliminate the Cauchy-Schwarz penalty. We note that this is not explicitly due to the cooperative nature of our approach; as shown in \ref{thm:non.cooperative.dmt} non-cooperative transmitters can get the full multiplexing gain using lattice codes. However, our cooperative approach {\em does} permit the transmitters to use the remaining power to secure rate and diversity gains.

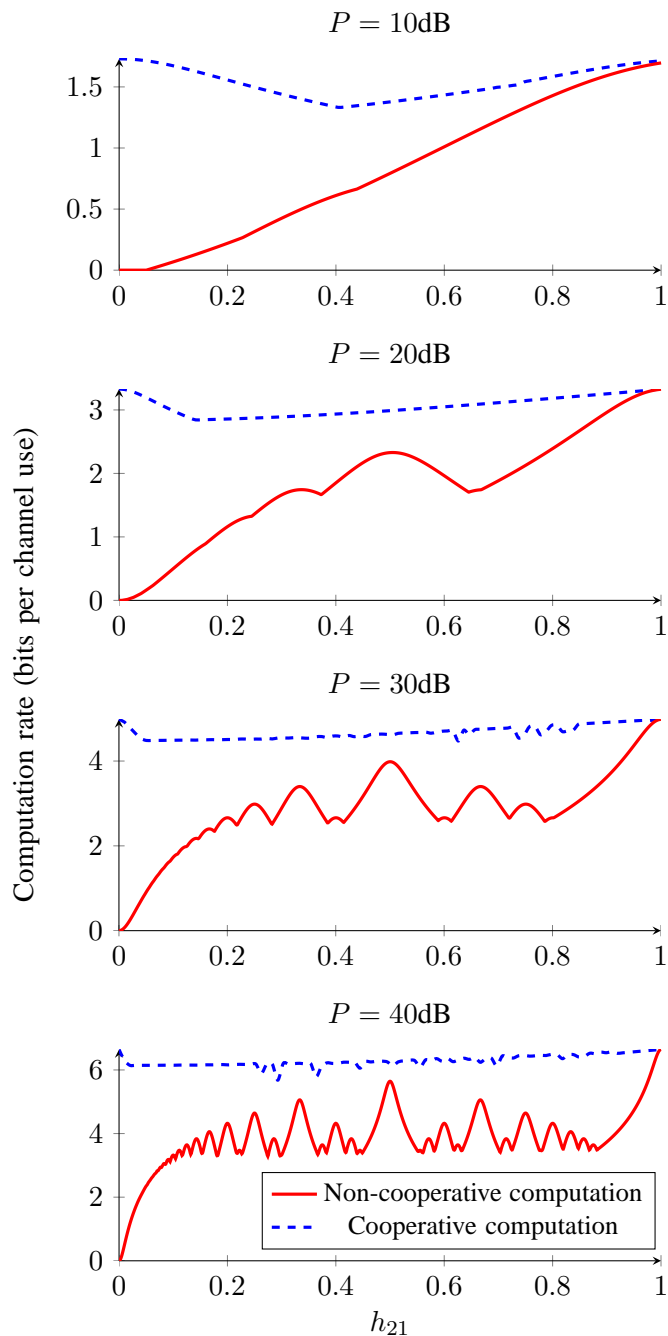
\begin{figure}[htb]
	\centering
\begin{tikzpicture}
	
	\begin{axis}[
	title={$P=10$dB},
	axis x line=bottom,
	axis y line=left,
	height=125pt,
	width=250pt
	]

		\addplot[smooth,color=blue,very thick,dashed] file {trans/cooperative.10.db};
		
		\addplot[smooth,color=red,very thick,solid] file {trans/noncooperative.10.db};
		

	\end{axis}
	
	\begin{scope}[yshift=-125pt]
	\begin{axis}[
	title={$P=20$dB},
	axis x line=bottom,
	axis y line=left,
	height=125pt,
	width=250pt
	]

		\addplot[smooth,color=blue,very thick,dashed] file {trans/cooperative.20.db};
		
		\addplot[smooth,color=red,very thick,solid] file {trans/noncooperative.20.db};
		

	\end{axis}
	\end{scope}
	
	\begin{scope}[yshift=-250pt]
	\begin{axis}[
	title={$P=30$dB},
	ylabel={\quad\quad\quad\quad\quad\quad\quad\quad\quad\quad\quad Computation rate (bits per channel use)},
	axis x line=bottom,
	axis y line=left,
	height=125pt,
	width=250pt
	]

		\addplot[smooth,color=blue,very thick,dashed] file {trans/cooperative.30.db};
		
		\addplot[smooth,color=red,very thick,solid] file {trans/noncooperative.30.db};
		

	\end{axis}
	\end{scope}
	
	\begin{scope}[yshift=-375pt]
	\begin{axis}[
	title={$P=40$dB},
	xlabel={$h_{21}$},
	legend style={at={(0.99,0.42)}},
	axis x line=bottom,
	axis y line=left,
	height=125pt,
	width=250pt
	]
		
		\addplot[smooth,color=red,very thick,solid] file {trans/noncooperative.40.db};
		\addlegendentry{\small Non-cooperative computation};

		\addplot[smooth,color=blue,very thick,dashed] file {trans/cooperative.40.db};
		\addlegendentry{\small Cooperative computation};
		

	\end{axis}
	\end{scope}

\end{tikzpicture}
	\caption{Achievable rates as a function of $h_{21}$ and $P$ for Example \ref{ex:two.by.one}.}
	\label{fig:two.by.one.rates}
\end{figure}
\end{example}

\begin{example}\label{ex:two.by.two}
Finally,  we examine the system depicted in Figure \ref{fig:two.by.two}. Here we have $L=M=2$, and again we set all channel gains to unity except for $h_{21}$. Again asymmetry prevents us from choosing $\mathbf{a}$ and the steering vectors easily. We iterate manually over the possible choices for $\mathbf{a}$, choose zero-forcing beamformers for $\mathbf{v}_1$ and $\mathbf{v}_2$, and numerically optimize over $\mathbf{v}_0$. In order for zero-forcing to succeed, we choose $\mathcal{B} = \{1,2\}$.

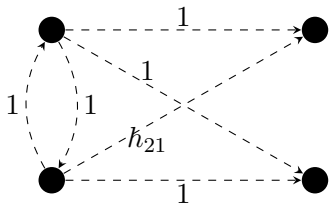
\begin{figure}[htb]
	\centering
\begin{tikzpicture}
		[scale=0.5,>=stealth,every node/.style={inner sep=0, minimum size=10}]
		\node[circle,fill] (x1) at (-2,2) {};
		\node[circle,fill] (x2) at (-2,-2) {}; 
				

		
		\draw [dashed,->] (x1) to [bend left]  node[right] {$1$} (x2);
		\draw [dashed,->] (x2) to [bend left] node[left] {$1$} (x1);


		\node[circle,fill] (y1) at (5,2) {};
		\node[circle,fill] (y2) at (5,-2) {};
		
		
		
		\draw [dashed,->] (x1) to node[above] {$1$} (y1);
		\draw [dashed,->] (x2) to node[below,pos=0.35] {$h_{21}$} (y1);
		
		\draw [dashed,->] (x1) to node[above,pos=0.35] {$1$}(y2);
		\draw [dashed,->] (x2) to node[below] {$1$} (y2);
		

\end{tikzpicture}
	\caption{Example \ref{ex:two.by.two}: A two-by-two computation network with asymmetric channel gains.}
	\label{fig:two.by.two}
\end{figure}

In Figure \ref{fig:two.by.two.rates} we plot the cooperative rate alongside (\ref{eqn:non.cooperative.rate}), again for a variety of signal-to-noise ratios. Again we make a few observations. In contrast to the previous scenario, here the rate drops when $h_{12} \approx 1$; this is because the channel matrix becomes increasingly ill-conditioned. Similar to before, in the cooperative case the rate remains non-zero, but here it occurs because the transmitters can cooperatively send a full-rank set of equations even though the channel matrix is nearly singular. However, in this example cooperation does not obtain the full multiplexing gain. The freedom to choose $\vbf_0$ allows us to mitigate the peakiness of the achievable rate, but we cannot eliminate the Cauchy-Schwarz penalty at both receivers simultaneously. Even for high SNR, however, we {\em do} get considerable robustness to channel variation.
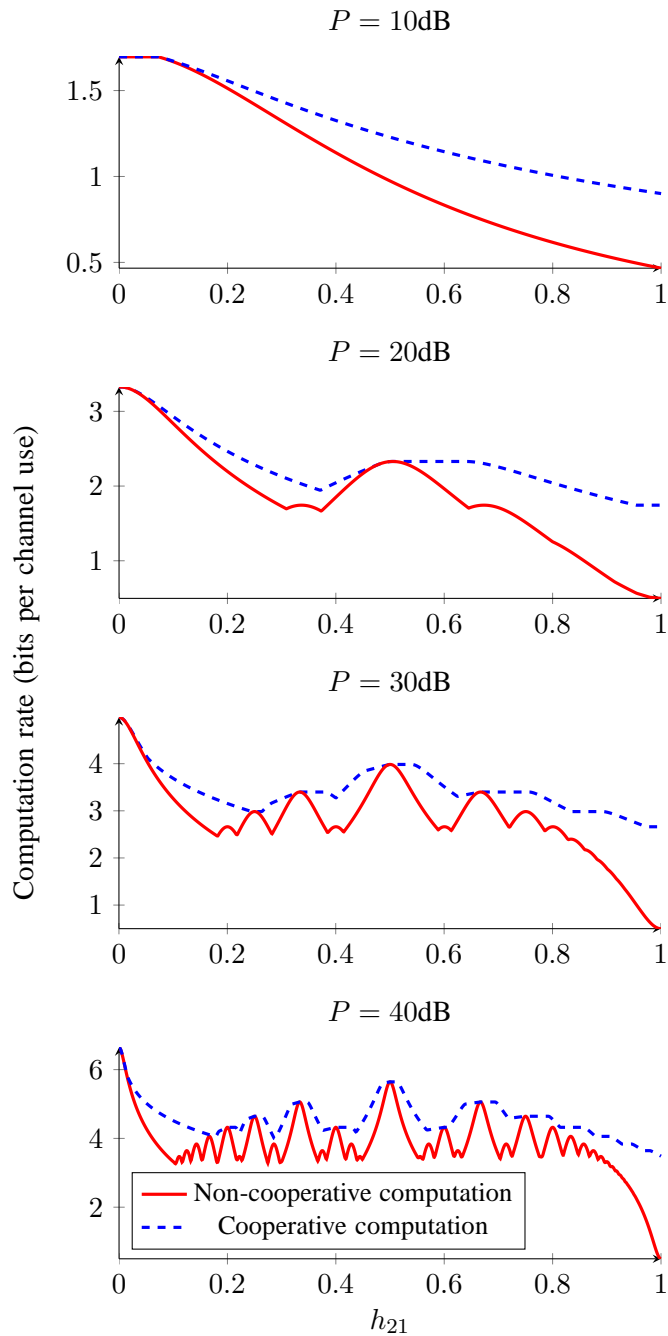
\begin{figure}[htb]
	\centering
\begin{tikzpicture}
	
	\begin{axis}[
	title={$P=10$dB},
	axis x line=bottom,
	axis y line=left,
	height=125pt,
	width=250pt
	]

		\addplot[smooth,color=red,very thick,solid] file {trans/noncooperative.2x2.10.db};
		
		\addplot[smooth,color=blue,very thick,dashed] file {trans/cooperative.2x2.10.db};
		

	\end{axis}
	
	\begin{scope}[yshift=-125pt]
	\begin{axis}[
	title={$P=20$dB},
	axis x line=bottom,
	axis y line=left,
	height=125pt,
	width=250pt
	]

		\addplot[smooth,color=blue,very thick,dashed] file {trans/cooperative.2x2.20.db};
		
		\addplot[smooth,color=red,very thick,solid] file {trans/noncooperative.2x2.20.db};
		

	\end{axis}
	\end{scope}
	
	\begin{scope}[yshift=-250pt]
	\begin{axis}[
	title={$P=30$dB},
	ylabel={\quad\quad\quad\quad\quad\quad\quad\quad\quad\quad\quad Computation rate (bits per channel use)},
	axis x line=bottom,
	axis y line=left,
	height=125pt,
	width=250pt
	]

		\addplot[smooth,color=blue,very thick,dashed] file {trans/cooperative.2x2.30.db};
		
		\addplot[smooth,color=red,very thick,solid] file {trans/noncooperative.2x2.30.db};
		

	\end{axis}
	\end{scope}
	
	\begin{scope}[yshift=-375pt]
	\begin{axis}[
	title={$P=40$dB},
	xlabel={$h_{21}$},
	legend style={at={(0.75,0.41)}},
	axis x line=bottom,
	axis y line=left,
	height=125pt,
	width=250pt
	]
		
		\addplot[smooth,color=red,very thick,solid] file {trans/noncooperative.2x2.40.db};
		\addlegendentry{\small Non-cooperative computation};
		
		\addplot[smooth,color=blue,very thick,dashed] file {trans/cooperative.2x2.40.db};
		\addlegendentry{\small Cooperative computation};
		

	\end{axis}
	\end{scope}

\end{tikzpicture}
	\caption{Achievable rates as a function of $h_{21}$ and $P$ for Example \ref{ex:two.by.two}.}
	\label{fig:two.by.two.rates}
\end{figure}
\end{example}


\section{Conclusion}\label{sect:conclusion}
We have studied the impact of user cooperation on compute-and-forward. Constructing a lattice-coding version of block Markov encoding, we presented a strategy that introduces a ``decode-and-forward'' element into computation coding. Transmitters decode each other's messages, enabling them to transmit resolution information cooperatively to the receivers. Our strategy achieves higher computation rates than previous approaches, since transmitters can jointly encode part of their messages, and coherent signals benefit from a beamforming gain. Additionally, cooperation enables an improvement in the diversity-multiplexing tradeoff, achieving full diversity when there is a single receiver.

In the case of multiple receivers, however, we have not established an achievable diversity-multiplexing tradeoff. The difficulty of aligning lattice codewords at multiple receivers suggests that lattice coding is insufficient for the task. A promising approach may be to introduce a cooperative element into the signal-alignment strategy of \cite{niesen:ISIT11}. Since this approach achieves the full multiplexing gain for multiple receivers, we expect to be able to obtain a non-trivial characterization of the DMT regardless of the number of receivers.

Finally, we note that our techniques can be applied to any situation in which one needs to merge lattice codes with decode-and-forward style cooperation. Our block Markov approach is rather general; as mentioned earlier, it can be used to achieve the capacity of the physically degraded relay channel or to achieve the decode-and-forward rates of the cooperative multiple-access channel. We therefore expect our techniques to be useful for developing new strategies and establishing new results in areas where lattice codes and cooperation are applied, such as physical-layer security \cite{lai:IT08,dong:SP10,ekrem:IT11,yuskel:IT11,huang:SP11} and interference channels \cite{sahin:globecom07,sridharan:ISIT08,rini:ITW10}.

\appendices
\section{Proofs of Upper Bounds}\label{app:upper.bounds}
Our first task is to prove Theorem \ref{thm:miso.upper.bound}, for which we need a quick lemma.
\begin{lemma}\label{lem:uniform.functions}
	Let $\wbf_1, \cdots, \wbf_L \in \mathbb{F}_p^k$ be independently and uniformly distributed messages. Then, the functions $\fbf_1, \cdots, \fbf_M$ are also independent and uniformly distributed across $\mathbb{F}_p^k$.
\end{lemma}
\begin{IEEEproof}
	Since the finite-field linear combinations in $\fbf_l$ are taken element-wise, it is sufficient to show the result for an arbitrary element of both messages and functions. Therefore, let $\wbf = (w_{11},\cdots,w_{L1})^T$ and $\fbf = (f_{11},\cdots,f_{M1})^T = \mathbf{A}\wbf$. We need to show that the elements of $\fbf$ are independent and uniformly distributed.
	
	Since $\wbf$ is uniformly distributed over $\mathbb{F}_p^L$, its probability mass function is
	\begin{equation}
		p(\wbf) = p^{-L}.
	\end{equation}
	The conditional pmf of $\fbf$ is
	\begin{equation}
		p(\fbf | \wbf) = \delta(\fbf - \mathbf{A}\wbf),
	\end{equation}
	where $\delta(\cdot)$ is the Kronecker delta function. Next we compute the marginal pmf for $\fbf$:
	\begin{align}
		p(\fbf) &= \sum_{\wbf \in \mathbb{F}_p^L}p(\fbf | \wbf)p(\wbf) \\
		&= p^{-L}\sum_{\wbf \in \mathbb{F}_p^L}\delta(\fbf - \mathbf{A}\wbf) \\
		&= p^{-L} \big | \left\{ \wbf | \mathbf{A}\wbf = \fbf\right\} \big | \label{eqn:full.rank} \\
		&= p^{-L}p^{L-M} = p^{-M},
	\end{align}
	where (\ref{eqn:full.rank}) follows because $\mathbf{A}$ is full rank. Since the pmf $p(\fbf)$ does not depend on $\fbf$, the elements are independent and uniformly distributed.
\end{IEEEproof}
With Lemma \ref{lem:uniform.functions}, it is straightforward to prove Theorem \ref{thm:miso.upper.bound}.
\begin{IEEEproof}[Proof of Theorem \ref{thm:miso.upper.bound}]
	Suppose that a genie provides the messages $\wbf_l(t)$ to each of the transmitters. Then the transmitters each can compute the functions $\fbf_m(t)$. By Lemma \ref{lem:uniform.functions} these functions are independent and uniformly distributed, the scenario is equivalent to an $L$-transmitter antenna having $M$ independent messages to send to $M$ users. In \cite{weingarten:IT06} the capacity region is shown to be (\ref{eqn:dirty.paper}). Since we define the computation capacity in terms of achievable {\em symmetric} rate, it cannot exceed the symmetric-rate MISO capacity given in (\ref{eqn:common.miso.rate}).
\end{IEEEproof}

Next we prove the upper bound in Theorem \ref{thm:relay.upper.bound}.
\begin{IEEEproof}[Proof of Theorem \ref{thm:relay.upper.bound}]
	Choose a transmitter $l$, and suppose that a genie supplies the messages $w_{l^\prime}(t)$ to the receivers for every $l^\prime \neq l$. By the crypto lemma \cite{erez:IT04}, each $\fbf_{m}(t)$ such that $a_{lm} \neq 0$ is statistically independent of the messages $w_{l^\prime}(t)$, so the receivers remain equivocal as to the desired functions. Thus the scenario is equivalent to a compound relay channel, with transmitter $l$ acting as the source, the transmitters $l^\prime$ acting as relays, and each receiver $m$ such that $a_{lm} \neq 0$ acting as destinations all needing the messages $w_{l}(t)$. The capacity of the compound relay channel can be bounded using cut-set arguments. For any cut $S \in \mathcal{S}_l$, the capacity of the compound relay channel, and thus the computation capacity of the cooperative compute-and-forward network, is bounded by
	\begin{align}
		C(\Hbf,\Gbf,P) &\leq \max_{p(\xbf)} \min_{m, a_{lm} \neq 0}  I(x_l,x_S; y_m, z_{S^C} | x_{S^C}) \\
		&\leq \min_{m, a_{lm} \neq 0} \max_{p(\xbf)} I(x_l,x_S; y_m, z_{S^C} | x_{S^C}).
	\end{align}
	Taking the minimum over all transmitters and all cuts $S$, we obtain the result.
\end{IEEEproof}

\section{Entropy of dithered lattices over the multiple-access channel}
Here we prove that the mutual information between dithered lattice codewords and any receiver approaches that of a Gaussian multiple-access channel.
\begin{lemma}\label{lem:lattice.to.gaussian}
	Let
	\begin{equation}
		\xbf_l = \sqrt{P}[\lambda_l + \tbf_l] \mod \Lambda_s
	\end{equation}
	be a collection of independent lattice codewords, dithered across the shaping lattice, for $1 \leq l \leq L$. Let	
	\begin{equation}
		\ybf = \sum_{l=1}^Lh_l \xbf_l + \nbf,
	\end{equation}
	be a noisy sum of the codewords, where the noise $\nbf$ has i.i.d. elements with variance $\sigma^2$. Then, for any set $\mathcal{B} \in \{1,\cdots,L\}$, the normalized mutual information between the transmit signals and the receive signal approaches at least that of a Gaussian multiple-access channel:
	\begin{equation}
		\lim_{n \to \infty}\frac{1}{n} I(\xbf_\mathcal{B} ; \ybf | \xbf_{\mathcal{B}^C}) \geq \frac{1}{2}\log_2\left(1+\frac{P\sum_{l \in \mathcal{B}}h_l^2}{\sigma^2}\right).
	\end{equation}
	When $\nbf$ is Gaussian, this bound is tight.
\end{lemma}
\begin{IEEEproof}
	Since $\ybf$ is the sum of transmitted signals, conditioning entails only subtracting away the known component. Therefore, letting
	\begin{equation}
		\ybf_\mathcal{B} = \sum_{l \in \mathcal{B}} h_l \xbf_l + \nbf,
	\end{equation}
	the mutual information is
	\begin{equation}
		\lim_{n \to \infty}\frac{1}{n} I(\xbf_\mathcal{B} ; \ybf | \xbf_{\mathcal{B}^C}) = \lim_{n \to \infty} \frac{1}{n}I(\xbf_\mathcal{B} ; \ybf_\mathcal{B}) = \lim_{n \to \infty}\frac{1}{n}(h(\ybf_{\mathcal{B}}) - h(\nbf)),
	\end{equation}	
	where $h(\cdot)$ is the differential entropy. Since the Gaussian distribution maximizes the differential entropy for a given variance, we have
	\begin{equation}
		\frac{1}{n}h(\nbf) \leq \frac{1}{2}\log(2\pi e \sigma^2). \label{eqn:n.entropy.bound}
	\end{equation}
	To bound $h(\ybf_\mathcal{B})$, we note that in \cite[Lemma 8]{nazer:IT11} it was shown that the density function $f_{\ybf_\mathcal{B}}$ is bounded by
	\begin{equation}
		f_{\ybf_{\mathcal{B}}} \leq e^{c(n)n}f_{\ybf^*},
	\end{equation}
	where $\ybf^*$ is an i.i.d. Gaussian vector with variance $P\sum_{l \in \mathcal{B}}h_l^2 + \sigma^2$, and $c(n)$ is a term approaching zero from above as $n \to \infty$. Plugging this into the definition of differential entropy, we have, for sufficiently high $n$,
	\begin{align}
		\frac{1}{n}h(\ybf_\mathcal{B}) &\geq -\frac{1}{n} \int e^{c(n)n}f_{\ybf^*} \log(e^{c(n)n}f_{\ybf^*}) \\
		&= -\frac{1}{n}e^{c(n)n}\int f_{\ybf^*} \log(f_{\ybf^*}) - \frac{1}{n} e^{c(n)n}c(n)n \\
		&= e^{c(n)n}\left(\frac{1}{n}h(\ybf^*) - c(n)\right) \\
		&\geq \frac{1}{n}h(\ybf^*) - c(n) \label{eqn:constant.is.small}\\
		&\to  \frac{1}{n}h(\ybf^*) \\
		&= \frac{1}{2}\log\left(2\pi e \left(P\sum_{l \in \mathcal{B}} + \sigma^2\right)\right), \label{eqn:y.entropy.bound}
	\end{align}
	where (\ref{eqn:constant.is.small}) follows because $e^{c(n)n} \geq 1$ and for sufficiently high $n$ the term $\frac{1}{n} h(\ybf^*) - c(n)$ is positive. Combining (\ref{eqn:n.entropy.bound}) and (\ref{eqn:y.entropy.bound}), we get that
	\begin{align}
		\lim_{n \to \infty}\frac{1}{n} I(\xbf_\mathcal{B} ; \ybf | \xbf_{\mathcal{B}^C}) &\geq \frac{1}{2}\log\left(2\pi e \left(P\sum_{l \in \mathcal{B}}h_l^2 + \sigma^2\right)\right) - \frac{1}{2}\log(2\pi e \sigma^2) \\
		&= \frac{1}{2}\log_2\left(1+\frac{P\sum_{l \in \mathcal{B}}h_l^2}{\sigma^2}\right).
	\end{align}
	When $\nbf$ is Gaussian, it is well-known that Gaussian inputs are optimal and result in the same mutual information as the bounds just established. In this case the bound is tight.
\end{IEEEproof}

\bibliography{/Users/nokleby/documents/LaTeX/bibliography}

\end{document}